\documentclass[acmtog]{acmart}
\acmSubmissionID{321}

\usepackage{booktabs} %

\usepackage[ruled]{algorithm2e} %

\SetAlFnt{\small}
\SetAlCapFnt{\small}
\SetAlCapNameFnt{\small}
\SetAlCapHSkip{0pt}

\usepackage{multirow}
\usepackage{makecell}
\usepackage{bbm}

\acmJournal{TOG}
\acmYear{2022}
\acmVolume{41}
\acmNumber{6}
\acmArticle{224}
\acmMonth{12}

\setcopyright{acmcopyright}

\acmDOI{10.1145/3550454.3555480}

\begin{document}
\title{Learning to Generate 3D Shapes from a Single Example}

\newcommand{\cmt}[1]{\marginpar{\Large{\color{red} $\spadesuit$}} {\bf \color{red} [#1]}}
\newcommand{\fixme}[1]{\textbf{\color{red}[#1]}}
\newcommand{\rev}[1]{{\color{blue}#1}} %

\newcommand{\secprespace}{\vspace{-2mm}}
\newcommand{\secspace}{\vspace{-2mm}}
\newcommand{\paraspace}{\vspace{-2mm}}

\newcommand{\figref}[1]{Fig.~\ref{fig:#1}}
\newcommand{\tabref}[1]{Table~\ref{tab:#1}}
\newcommand{\secref}[1]{Sec.~\ref{sec:#1}}
\newcommand{\thmref}[1]{Theorem~\ref{thm:#1}}
\newcommand{\lemref}[1]{Lemma~\ref{lem:#1}}
\newcommand{\appref}[1]{Appendix~\ref{sec:#1}}
\newcommand{\algref}[1]{Algorithm~\ref{alg:#1}}
\newcommand{\eq}[1]{\eqref{eq:#1}}
\newcommand{\norm}[1]{\left\lVert#1\right\rVert}

\newcommand{\txt}[1]{\textrm{#1}}
\newcommand{\bb}[0]{\bm{b}}
\newcommand{\sfd}[0]{\mathsf{d}}
\newcommand{\sfW}[0]{\mathsf{W}}
\newcommand{\kwta}[0]{$k$-WTA\xspace}
\newcommand{\x}[0]{\mathbf{x}}

\makeatletter
\DeclareRobustCommand\onedot{\futurelet\@let@token\@onedot}
\def\@onedot{\ifx\@let@token.\else.\null\fi\xspace}

\def\eg{\emph{e.g}\onedot} \def\Eg{\emph{E.g}\onedot}
\def\ie{\emph{i.e}\onedot} \def\Ie{\emph{I.e}\onedot}
\def\cf{\emph{c.f}\onedot} \def\Cf{\emph{C.f}\onedot}
\def\etc{\emph{etc}\onedot} \def\vs{\emph{vs}\onedot}
\def\wrt{w.r.t\onedot} \def\dof{d.o.f\onedot}
\def\etal{\emph{et al}\onedot}
\makeatother
\newcommand{\rw}[1]{{\color{red}\{\textbf{Rundi:} #1\}}}
\newcommand{\cz}[1]{{\color{green}\{\textbf{Changxi:} #1\}}}

\newcommand{\Ladv}{\mathcal{L}_\text{adv}}
\newcommand{\Lrec}{\mathcal{L}_\text{rec}}

\newcommand{\upsample}{\uparrow}
\newcommand{\xrec}[1]{\tilde x_{#1}^*}
\author{Rundi Wu}
\affiliation{%
  \institution{Columbia University}
  \city{New York City}
  \state{NY}
  \postcode{10025}
  \country{USA}}
\email{rundi.wu@columbia.edu}

\author{Changxi Zheng}
\affiliation{%
  \institution{Columbia University}
  \city{New York City}
  \state{NY}
  \postcode{10025}
  \country{USA}}
\email{cxz@cs.columbia.edu}

\begin{abstract}
Existing generative models for 3D shapes are typically trained on a large 3D
dataset, often of a specific object category.  In this paper, we investigate
the deep generative model that learns from only a \emph{single} reference 3D shape.
Specifically, we present a multi-scale GAN-based model designed to
capture the input shape's geometric features across a range of spatial scales.  
To avoid large memory and computational cost induced by operating on the 3D volume, 
we build our generator atop the
tri-plane hybrid representation, which requires only 2D
convolutions.  We train our generative model on a voxel pyramid of the
reference shape, without the need of any external supervision or manual
annotation.  Once trained, our model can generate diverse and high-quality 3D
shapes possibly of different sizes and aspect ratios. The resulting shapes
present variations across different scales, and at the same time retain the
global structure of the reference shape. Through extensive evaluation, both
qualitative and quantitative, we demonstrate that our model can generate
3D shapes of various types.\footnote{Project webpage: \url{http://www.cs.columbia.edu/cg/SingleShapeGen/}}
\end{abstract}



\begin{teaserfigure}
    \includegraphics[width=\linewidth]{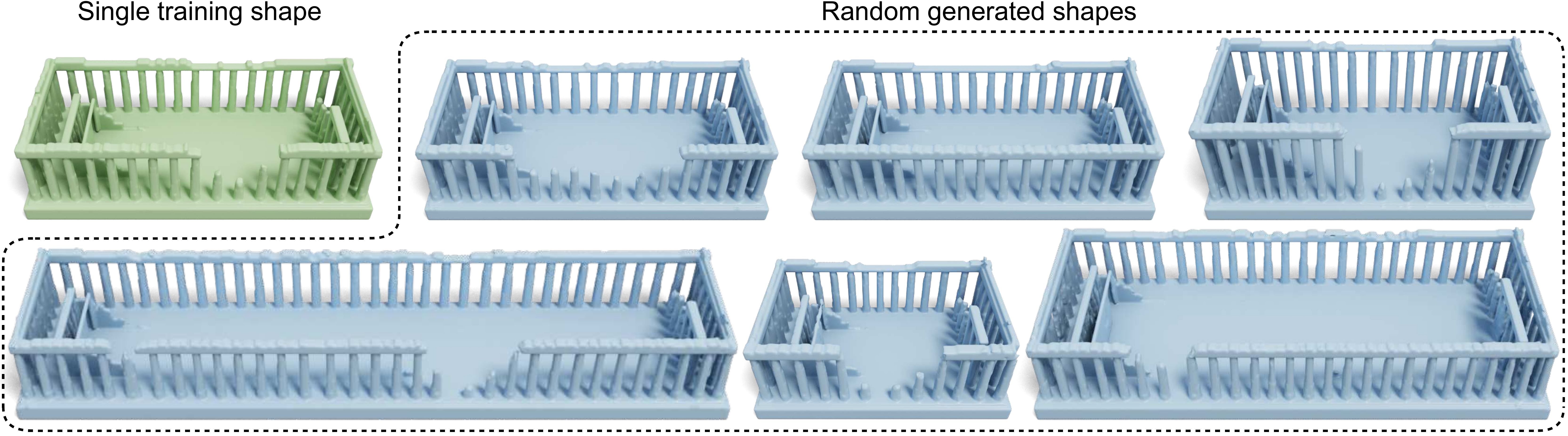}
    \caption{\textbf{A gallery of acropolises.} Trained on a single acropolis shape (top left), our proposed
        generative model is able to produce a diverse gallery of acropolises, possibly
        of different sizes and aspect ratios. The generated shapes depict rich
        variations (such as the dents and breakage of the columns) across different scales,
        and at the same time retain the essential structure of the reference shape (such as the layout of the columns on the rectangular base).
        \copyright The original 3D acropolis model (top left) by choly kurd under Standard License Editorial Use Only (turbosquid.com).
    }
    \label{fig:teaser}
\end{teaserfigure}

\maketitle

\section{Introduction}
The user creation of novel 3D digital shapes is nontrivial, 
requiring technical know-how and a sense of art, often taking time and patience.
It is such a laborious process that motivates the development of computer algorithms
that create shapes\textemdash 
new, diverse, and high-quality 3D shapes created in a fully automatic fashion.
Leveraging recent advance in deep learning, research in this direction has been vibrant~\cite{wu2016learning,achlioptas2018learning,chen2019learning,nash2020polygen,jayaraman2022solidgen}:
the general theme here is to develop a generative model able to learn from a training dataset to generate 3D shapes.

The training dataset for almost all existing 3D generative models must be sufficiently large\textemdash 
often provided in a specific category such as chairs~\cite{shapenet2015}, mechanical
parts~\cite{willis2021fusion} and terrains~\cite{guerin2017interactive}.
However, collecting a large set of 3D shapes in the first place is by no means an easy task.
Unlike images, which can be easily captured by cameras and are widely available online,
3D shapes require significant manual effort to 3D scan or model in modern 3D modeling software.
Therefore, to use a learning-based generative model for creating 3D shapes, 
one must address the chicken and egg problem, unless the generative model is 
liberated from demanding a large training dataset.

Our work aims for this liberation.
We present a deep generative model that learns from just a single 3D shape,
without the need of any manual annotation or external data.
Our goal is conceptually similar to example-based 2D texture synthesis~\cite{wei2009state}: 
produce new, diverse, as many as needed samples from a given input example.
But our model is not merely meant to generate stationary 3D textures.  It strives
to produce diverse shape variations while preserving the global structure
presented in the input shape.

As an example shown in~\figref{teaser}, an acropolis shape is provided to train
our generative model. The generated shapes all share the same global structure as
the input example: a number of columns are placed on the edge of a rectangular base, leaving the central chamber area open. 
Meanwhile, they are all different. They have varying local features, such as different dents and breakage of the columns.
Our model can further synthesize new shapes that have bounding box sizes and aspect ratios different from the input.
We refer the reader to ~\figref{gallery} and the appendix (\figref{gallery2} and \figref{gallery3}) for 
a range of example shapes and synthesized results.

\begin{figure*}
  \centering
  \includegraphics[width=0.98\textwidth]{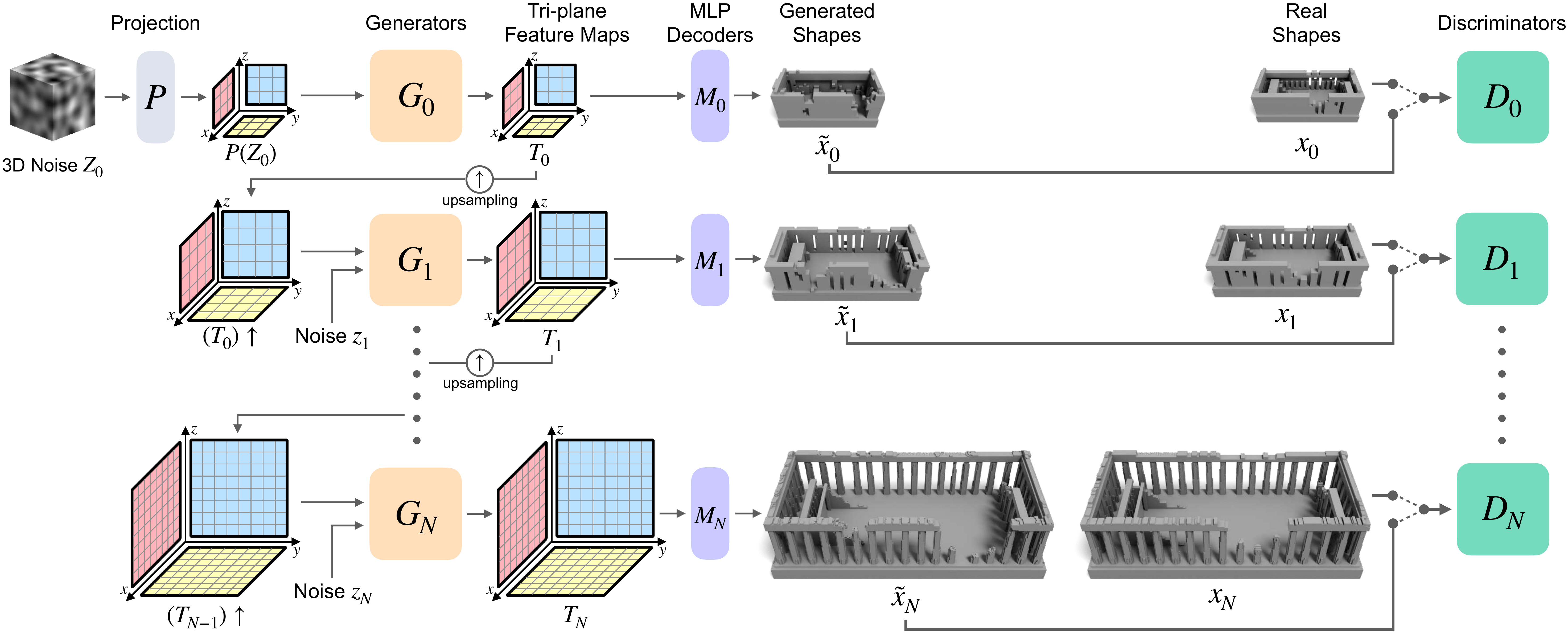}
  \vspace{-1mm}
  \caption{\textbf{Model overview.} 
  	Our model consists of a pyramid of generators $\{G_i\}$ and discriminators $\{D_i\}$, trained on a voxel pyramid $\{x_i\}$ of a single reference shape.
  	We train our model in a progressive manner, from the coarsest scale (top) to the finest scale (bottom).
        The shape generation starts with a random 3D noise $Z_0$ of same dimensions as the
        coarsest voxel grid $x_0$. $Z_0$ is first projected by a network $P$
        on the three feature planes to obtain the initial tri-plane feature map $P(Z_0)$,
        which is then fed into the first generator $G_0$.
  	The generator $G_i$ at level $i$ outputs a tri-plane map $T_i$, 
        which is subsequently upsampled and served as input to the next level. 
        When training the $i$-th level discriminator $D_i$, the output tri-plane feature map $T_i$
        is decoded by an MLP decoder $M_i$ to get the generated shape $\tilde x_i$ (in a voxelized representation). 
        $D_i$ aims to distinguish individual patches in $\tilde x_i$ from their counterparts in $x_i$.  
        More details are presented in ~\secref{msgen}.}
  \label{fig:framework}
  \vspace{-1mm}
\end{figure*}

Technically, our work is inspired by the recent advance in single image generative adversarial networks (GANs)~\cite{shaham2019singan,hinz2021improved},
wherein the goal is to learn the distribution of image patches on a single input image.
Similar in spirit to those works, our generative model is based on a multi-scale, hierarchical GAN architecture (see \figref{framework}), 
trained {on} a voxel pyramid of the input 3D shape.
The voxel pyramid is responsible for capturing geometric features across multiple scales, from global structures to local details.
However, since each level of the voxel pyramid is a 3D grid, it has a large memory footprint.
To make the matter worse, the 3D convolution needed to operate on a 3D grid produces intermediate feature maps that require even larger memory.
The intensive memory requirement severely limits the grid resolution and in turn the generative model's ability to learn geometric details.

We therefore seek for sidestepping 3D convolutions. Our generative model operates on the tri-plane feature map\textemdash which 
encodes a shape in three axis-aligned 2D feature maps\textemdash
followed by a small multilayer perceptron (MLP) network that describes the generated shape as a neural 
implicit function \cite{Mescheder_2019_CVPR,peng2020convolutional}. 
The use of tri-plane feature map significantly reduces memory and computation cost. 
It allows the generative model to learn shape features across a wide range of scales,
and the MLP network enables the model to output a shape at an arbitrary resolution.

To our knowledge, this is the first deep generative model that synthesizes novel 3D shapes from a single example while
capturing shape features across multiple scales.
We demonstrate generation results on various 3D shapes of different categories.
We also perform quantitative evaluations to compare our method to several baselines and prior methods.
In addition, we provide ablation studies to justify our network design, training strategy, and data construction.

\section{Related Work}

\paragraph{3D model synthesis}
Similar to texture synthesis~\cite{wei2009state}, traditional
3D shape synthesis techniques can be largely classified into 
procedural and example-based methods.  Procedural modeling techniques~\cite{ebert2003texturing} 
have been extensively studied over the years.  They
typically require the specification of many rules for generating shapes,
and therefore they are often designed for specific classes of objects, such as terrains
\cite{musgrave1989synthesis,smelik2009survey}, cities
\cite{parish2001procedural,muller2006procedural,talton2011metropolis}, and
trees \cite{mvech1996visual,prusinkiewicz2001use,longay2012treesketch}.

Example-based methods are more general-purpose, aiming to synthesize new shapes by analyzing the given example(s).
The pioneering work in this direction~\cite{funkhouser2004modeling} and others~\cite{kalogerakis2012probabilistic,xu2012fit} 
synthesize novel shapes
by assembling components retrieved from a database of segmented shapes.
Another line of works
\cite{merrell2007example,merrell2008continuous,zhou2007terrain,bokeloh2010connection}
analyze a single input shape and generate larger 3D models by exploiting
repeated patterns within the input example.  However, these techniques either
rely on manual decomposition \cite{merrell2007example} or require exact
symmetry \cite{bokeloh2010connection}.

Our work also falls into the category of example-based 3D modeling. In
contrast to prior methods that usually rely on hand-crafted rules, we
leverage neural networks to automatically learn multi-scale shape
features of the single input example.  Our method is also related to
multi-scale texture synthesis~\cite{han2008multiscale}, as we build our network
in a hierarchical manner and train it on a voxel pyramid constructed from the
input 3D shape.

\paragraph{Deep generative models for 3D shapes.}
Since the introduction of deep generative networks such as Generative Adversarial Network (GAN) \cite{Ian2014gan} and Variational Autoencoder (VAE) \cite{Kingma2014vae}, 
developing deep generative models for 3D shapes has attracted immense research interest.
Existing works learn to produce 3D shapes in different representations,
including voxels \cite{wu2016learning,chen2021decor}, point clouds \cite{achlioptas2018learning,yang2019pointflow,cai2020learning,li2021sp}, meshes \cite{nash2020polygen,Hertz2020deep,liu2020neural,gao2021tm,pavllo2021learning}, implicit functions \cite{chen2019learning,park2019deepsdf,Mescheder_2019_CVPR,Kleineberg2020}, multi-charts \cite{ben2018multi,groueix2018papier},structural primitives \cite{li2017grass,mo2019structurenet,wu2020pq,jones2020shapeAssembly}, and parametric models \cite{chen2020bsp,wu2021deepcad,jayaraman2022solidgen}.
Nearly all these methods are trained on a large dataset of category-specific 3D shapes, \eg, ShapeNet \cite{shapenet2015}.

Although the use of a large dataset enables the generative model to learn 
rich information shared across various shape instances, 
it also limits the application scope of these methods.
Unlike images, which can be easily captured and downloaded, collecting a large,
high-quality dataset of 3D shapes is much more expensive.  Moreover, many
artistically designed shapes have unique structures, 
and thus it is hard, if not impossible, to learn them from a large data collection.
In this work, instead of learning a shape distribution from a large dataset, we focus on 
learning shape features across multiple scales from a single 3D shape. 
Without the need of a large 3D dataset, 
even from a unique shape (\eg, designed by a 3D artist),
our method is able to learn and generate similar but new shapes.

\paragraph{Learning a generative model from a single example}
In several recent works, researchers start to explore generative models that
are learned from a single example in various data domain, 
such as
images~\cite{zhou2018non,shaham2019singan,hinz2021improved,shocher2019ingan,granot2021drop,chen2022exemplar},
videos~\cite{haim2021diverse}, audio~\cite{greshler2021catch}, 3D
meshes~\cite{Hertz2020deep} and motion sequences~\cite{li2022ganimator}.
There are also data-driven methods on 3D mesh reconstruction~\cite{Hanocka2020p2m} and stylization~\cite{michel2022text2mesh} under one-shot setting.
A common theme behind these works is to capture the internal patch
distribution of a given example, and then apply the learned knowledge in 
the generation process.
The seminal work, SinGAN~\cite{shaham2019singan},
proposes to learn a single image generative model by training a hierarchy
of 2D convolutional patch-GANs.  Their method is not restricted to stationary texture synthesis,
and it can also synthesize natural images that present large-scale structures.  
But one can not simply transfer their method for 3D shape synthesis, as it would require
3D CNNs on 3D voxel grids, which are inefficient and memory intensive, 
especially when a high resolution output is desired.

To the best of our knowledge, DGTS~\cite{Hertz2020deep} is the only prior work
on learning a generative model from a single 3D shape.  It learns a
convolutional network on triangle meshes, 
predicting vertex displacement after subdivision.  
However, this method is mainly designed for geometric texture transfer, thus limited to 
the synthesis of stationary and isometric local geometric features.  
Moreover, its generated shapes are limited to the topology of the input.  
Our proposed method, on the contrary, learns from a voxel representation. 
It can capture geometric features across
multiple scales and generate samples that have different topologies from the input.
To avoid the memory cost for directly operating on 3D voxel grids, 
our generator leverages an efficient tri-plane hybrid representation, 
and thereby only 2D convolutions are needed.

\section{Method}\label{sec:method}
\paragraph{Overview}
A natural way of capturing shape features across multiple scales is to use a
hierarchical representation of the input shape. Provided an input shape $x$, we
{voxelize it} and construct a voxel pyramid $\{x_0, ..., x_N\}$, where $x_i$
indicates a voxelized version of $x$ at resolution $s_N\cdot r^{N-i}$, for a
user-specified size $s_N$ and a downsampling factor $r<1$.
All $x_i$ are voxelized independently. 
Correspondingly, our generative model consists of a hierarchy of GANs (see our
network overview in \figref{framework}).
At each level $i$, the generator aims to synthesize a shape indistinguishable from $x_i$.
Its output is fed as an input to the generator at level $i+1$ to propagate the synthesized large-scale
structure to the finer level.

Crucial to this hierarchy of GANs is the structure of each generator.
To operate on voxel grids, typically used is the 3D convolution.
But the hierarchy of voxel grids and the 3D convolution results\textemdash which 
extend a 3D grid by another dimension, namely the number of channels\textemdash
require a significant amount of memory (see \tabref{cost} later).  
This intensive memory footprint prevents us from
using a high resolution grid, thus limiting the finest features that the generative model can learn. 
To overcome this limitation, we design the generator 
to operate \emph{not} on a voxel grid directly, but on its tri-plane feature map (see \secref{triplane}). 
The generator takes as input a tri-plane map and outputs another tri-plane map.
The output map, equipped with an MLP network, serves as
a neural implicit function\textemdash one that allows the output shape to be constructed at an arbitrary resolution.

The discriminators in our GAN hierarchy have simple structures,
each composed of three convolutional layers. 
A discriminator is responsible for distinguishing the synthesized and input shape in terms of
local voxel patches of the same size ($11\times 11 \times 11$ in our implementation). 
On a different level of the hierarchy, the same size of a local patch captures shape features in different scales.
In this way, the low-level discriminators force our generative model to preserve 
large-scale global structures, while the high-level discriminators encourage small-scale local variations.

\subsection{Tri-plane Hybrid Representation}
\label{sec:triplane}

\begin{figure}[t]
  \centering
  \includegraphics[width=0.94\linewidth]{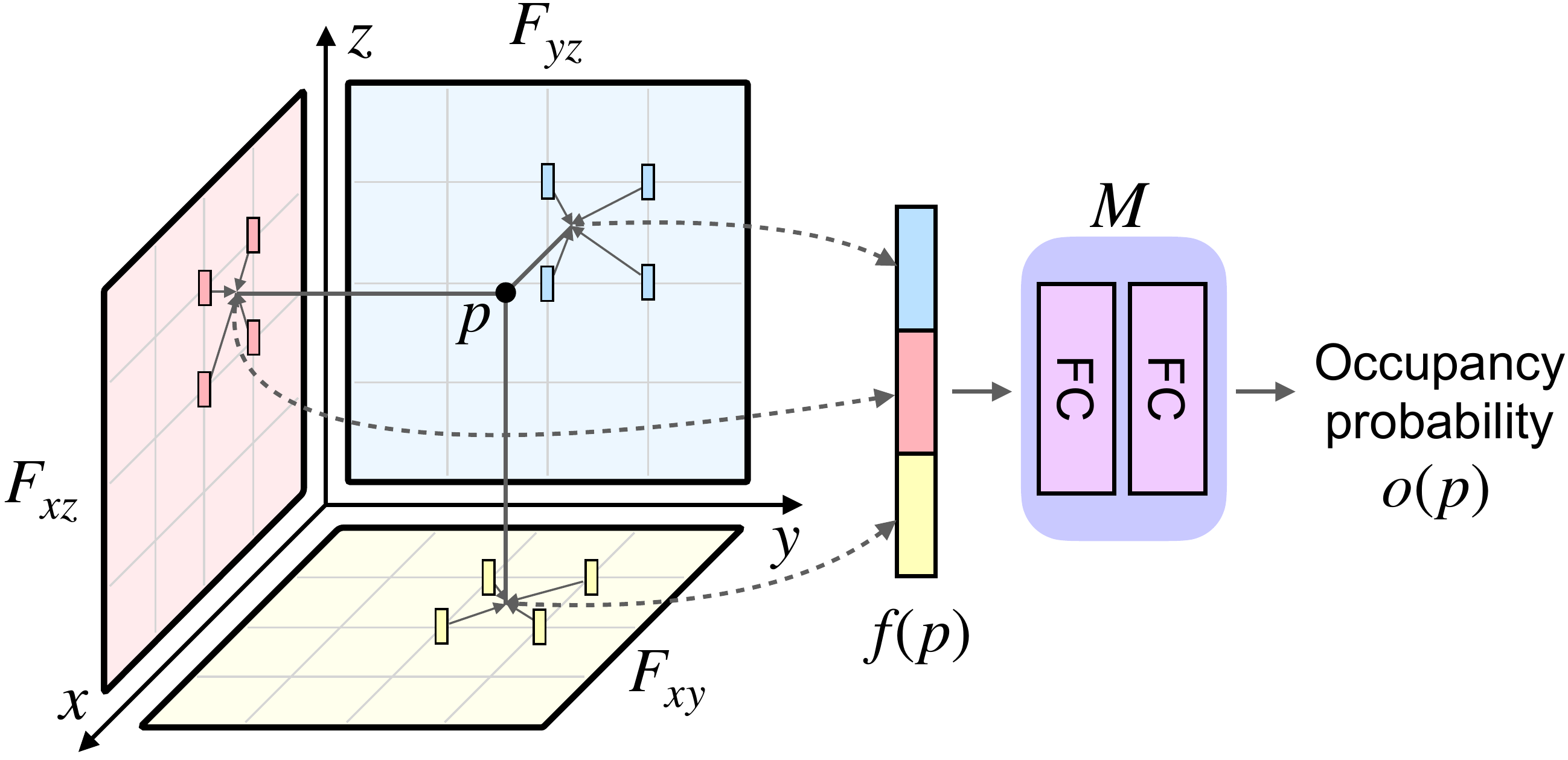}
  \vspace{-2mm}
  \caption{\textbf{Tri-plane feature map.} A spatial position $p$ is projected to
  three feature planes $(F_{xy}, F_{xz}, F_{yz})$ to query the corresponding
  features, which are then concatenated and fed into an MLP $M$ to obtain 
  $o(p)$, the probability of $p$ being occupied by the 3D shape.
  Here for the sake of visualization, the three planes have equal size. In
  practice, they can be of different aspect ratios as long as their dimensions
  agree with each other (See~\secref{triplane} for details).}
 \label{fig:triplane}
 \vspace{-2mm}
\end{figure}

Although a 3D shape has a volume, its shape features are on its
surface; away from its surface, the grid data is featureless.
We therefore seek to encode a 3D shape into 2D feature maps.
In particular, we build our generative model upon the tri-plane
hybrid representation emerged in recent neural implicit function literature~\cite{peng2020convolutional,chan2021efficient,Chen2022ECCV}.

Instead of representing a 3D shape on a voxel grid of size $D\times H\times W$, 
the tri-plane hybrid representation expresses the 3D shape using 
a tri-plane feature map $T$ together with a small MLP network $M$.
The tri-plane feature map $T$ contains three axis-aligned 2D feature maps,
\begin{equation}
	T=(F_{xy}, F_{xz}, F_{yz}), 
\end{equation}
where $F_{xy}\in \mathbb{R}^{C\times D\times H}$, $F_{xz}\in
\mathbb{R}^{C\times D\times W}$ and $F_{yz}\in \mathbb{R}^{C\times H\times W}$,
with $C$ being the number of channels.  
Consider a 3D position $p\in \mathbb{R}^3$. Its occupancy by a 3D shape is related to its tri-plane feature map 
in the following way (see \figref{triplane}).
First, $p$ is projected onto the three axis-aligned planes to obtain its 2D coordinates,
$p_{xy}$, $p_{xz}$, and $p_{yz}$, respectively. Based on the 2D coordinates, we query the tri-plane feature map
to obtain $p$'s feature vector $f(p)$:
\begin{equation}\label{eq:tri_plane}
\begin{split}
f_{xy}(p) &= \text{interp}(F_{xy}, p_{xy}), \\
f_{xz}(p) &= \text{interp}(F_{xz}, p_{xz}), \\
f_{yz}(p) &= \text{interp}(F_{yz}, p_{yz}), \\
f(p) &= [f_{xy}, f_{xz}, f_{yz}],
\end{split}
\end{equation}
where $\text{interp}(\cdot, q)$ performs bilinear interpolation of a 2D feature map at position $q$,
akin to a 2D texture value lookup; and $f(p)\in\mathbb{R}^{3C}$ is simply a concatenation of the three interpolation results.
Lastly, an additional lightweight MLP decoder $M$ takes in $f(p)$ and outputs $p$'s occupancy probability,
\begin{equation}\label{eq:mlp_decoder}
	o(p)=M(f(p)),
\end{equation}
which is an implicit representation of the 3D shape. If $o(p)$ is determined, one can use the Marching Cubes algorithm~\cite{marchingcubes}
to construct the shape's triangle mesh.

Note that unlike previous works~\cite{peng2020convolutional,chan2021efficient}, 
here we intentionally exclude the 3D position of $p$ from the input of the MLP.
This is to ensure the generator position-invariant\textemdash a desired property that will become 
clear in \secref{msgen}.

From now on, we denote $M(T)\in\mathbb{R}^{D\times H\times W}$ as decoding the
entire 3D volume of grid size $D\times H\times W$ from the tri-plane feature map $T$, 
a shorthand for querying every grid point through $M$.

\begin{figure}[t]
  \centering
  \includegraphics[width=0.99\linewidth]{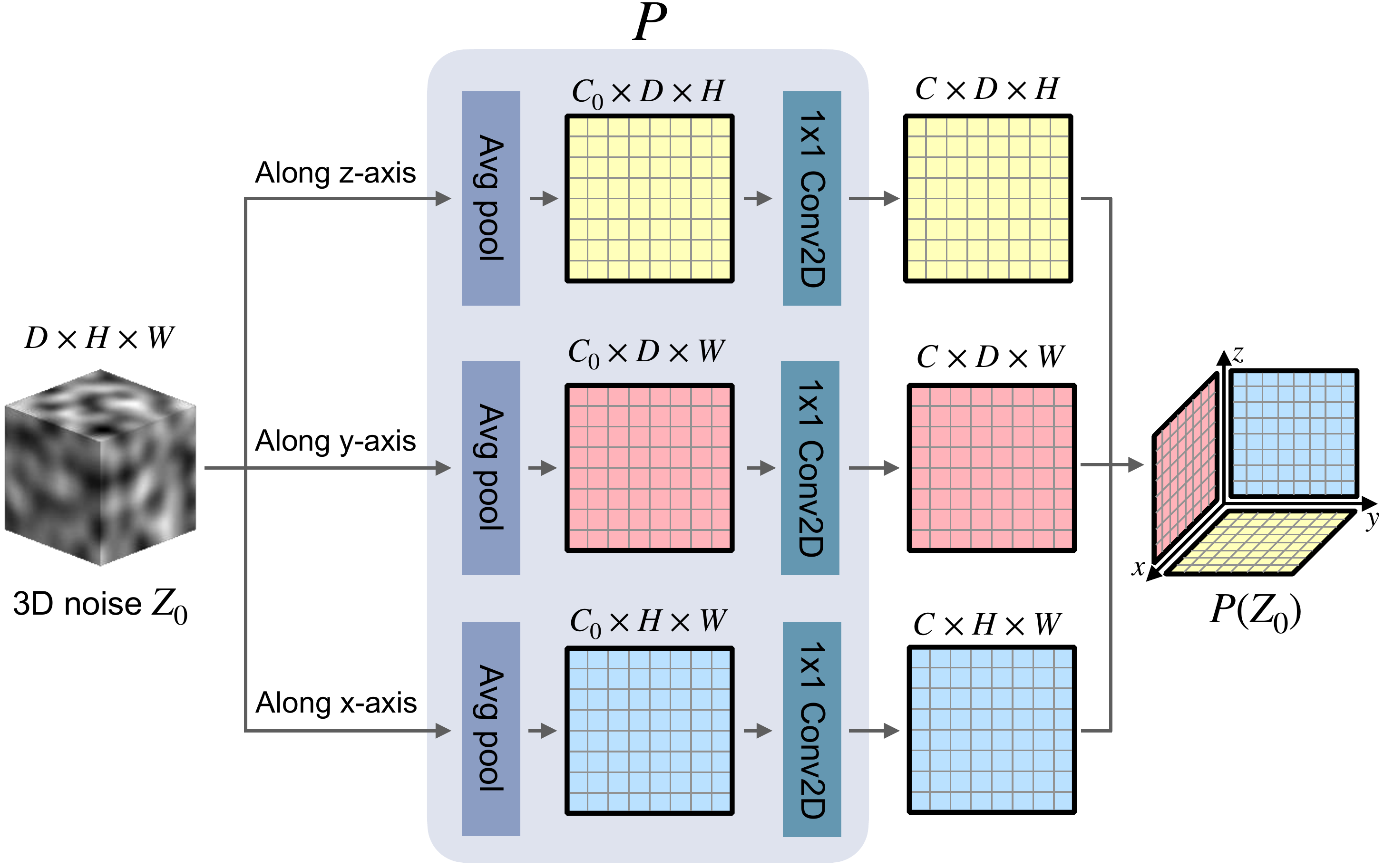}
  \vspace{-2mm}
  \caption{\textbf{The projection network $\mathbf{P}$.} Random 3D noise $Z_0$
  is projected by $P$ along three axes onto three axis-aligned feature planes. Along 
  each axis, $Z_0$ is first average pooled to a 2D feature map with $C_0$
  channels, which is then lifted to $C$ channels through a 2D convolutional
  layer with kernel size $1\times 1$. }
  \label{fig:projection}
\end{figure}

\subsection{Multi-scale Generation}\label{sec:msgen}
For the purpose of multi-scale shape generation, we need to organize tri-plane
feature maps in a hierarchical fashion.  This also sets our tri-plane maps
apart from those previously used in neural implicit
functions~\cite{peng2020convolutional,chan2021efficient}. 

\paragraph{Generator}
In our GAN hierarchy, the generator $G_i$ at level $i$ outputs a tri-plane feature map $T_i$ whose
spatial resolution matches $x_{i}$ (see \figref{framework}). The output $T_i$
on the one hand is decoded by an MLP network $M_i$ into a generated shape $\tilde x_i=M_i(T_i)$,
which is in turn examined by the discriminator $D_i$ at the same level.
On the other hand, $T_i$, after upsampled, serves as an input to the generator $G_{i+1}$ at the next level to produce tri-plane map $T_{i+1}$, namely,
\begin{equation}
	T_{i+1} =G_{i+1}((T_{i})\upsample, z_{i+1}),\qquad i\ge 0,
\end{equation}
where $(T_{i})\upsample$ bilinearly upsamples each of the three 2D feature maps
$(F_{xy}^i, F_{xz}^i, F_{yz}^i)$ by a factor of $\frac{1}{r}$, such that their
spatial dimensions match $x_{i+1}$.
In this way, the larger scale shape structures synthesized by $G_i$ is propagated into 
the next generator $G_{i+1}$ to add finer scale details.
The second input $z_{i+1}$ indicates the noise to introduce randomness in the generation process,
containing three independent spatial Gaussian noise maps, $z_{xy}^{i+1}$, $z_{xz}^{i+1}$ and  $z_{yz}^{i+1}$.

\begin{figure}[t]
  \centering
  \includegraphics[width=0.99\linewidth]{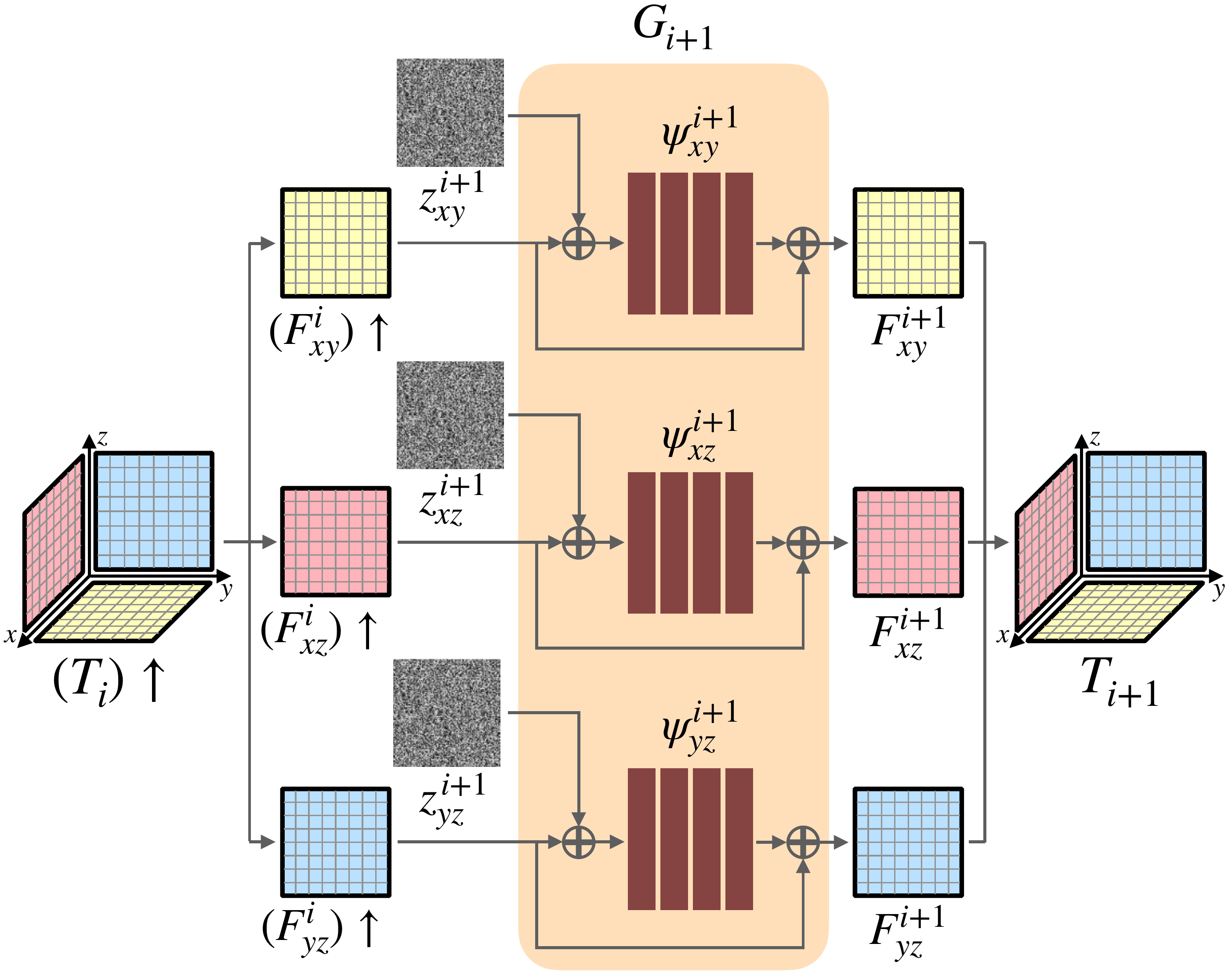}
  \vspace{-2mm}
  \caption{\textbf{Generator structure}. To refine the upsampled tri-plane feature
  map $(T_{i})\uparrow^\frac{1}{r}=((F_{xy}^i)\upsample, (F_{xz}^i)\upsample,
  (F_{yz}^i)\upsample)$ from the previous scale $i$, we first add three
  independent noise maps $(z_{xy}^{i+1},z_{xz}^{i+1},z_{yz}^{i+1})$.  They are
  then fed into three separate 2D convolutional nets
  ($\psi_{xy},\psi_{xz},\psi_{yz}$), each containing $4$ convolutional blocks
  with kernel size $3\times 3$.  There is a skip connection in each net that
  adds the input feature maps to its output, producing the final output of
  $G_{i+1}$, \ie, $T_{i+1}=(F_{xy}^{i+1}, F_{xz}^{i+1}, F_{yz}^{i+1})$.
  }\label{fig:generator}
  \vspace{-1mm}
\end{figure}

Concretely, the generator $G_{i+1}$ performs the following operations,
\begin{equation}
\begin{split}
F_{xy}^{i+1}=&(F_{xy}^i)\upsample+\psi_{xy}^{i+1}(z_{xy}^{i+1}+(F_{xy}^i)\upsample), \\
F_{xz}^{i+1}=&(F_{xz}^i)\upsample+\psi_{xz}^{i+1}(z_{xz}^{i+1}+(F_{xz}^i)\upsample), \\
F_{yz}^{i+1}=&(F_{yz}^i)\upsample+\psi_{yz}^{i+1}(z_{yz}^{i+1}+(F_{yz}^i)\upsample), \\
\end{split}
\end{equation}
where $\psi_{xy},\psi_{xz},\psi_{yz}$ are three 4-layer 2D convolutional networks,
wherein each layer has a kernel size $3\times 3$.
Figure~\ref{fig:generator} depicts the generator structure for $i \ge 1$.

The first generator $G_0$ is slightly different.
Although it has the same network structure as others, it takes only one input, the tri-plane map,
but not the noise. Thus, it has no skip connections (those shown in \figref{generator}).
Since there is no lower level generator to provide the input tri-plane map, 
the tri-plane map to $G_0$ comes from a 3D noise $Z_0$: %
We use a network module $P$ (see \figref{projection} for its structure)
to project $Z_0$ and produce the initial tri-plane feature map $P(Z_0)$. In short, the first 
generator $G_0$ produces a tri-plane map $T_0$ through $T_0=G_0(P(Z_0))$.

We note that a seemingly simpler option is to use three independent 2D noise maps to form a 
tri-plane map input to $G_0$, and thereby no projection network $P$ is needed. 
This approach, however, produces suboptimal results, because according to the
tri-plane map construction in~\eq{tri_plane}, its three 2D feature maps must be
correlated. See \secref{ablation} for an ablation study on our choice.

Overall, the generators in our GAN hierarchy are fully convolutional, operating solely on 2D feature maps. 
Also the MLP decoder in \eq{mlp_decoder} is position-invariant. Consequently, at inference time, we are able to generate shapes
with the sizes and aspect ratios different from the input shape (see~\figref{gallery}). 
This can be easily achieved by sampling an input 3D noise $Z_0$ with user specified spatial dimensions,
not necessarily the input shape's spatial dimensions. 

\paragraph{Discriminator}
The discriminators in our GAN hierarchy take as input either the generated
shape $\tilde x_i = M_i(T_i)$ or the example shape $x_i$, both of which are
provided in voxelized representation and have the same resolution.
The discriminators all have the same simple structure:
a 3-layer 3D convolutional network, wherein each layer has the kernel size $3\times 3 \times 3$, 
and the first layer (with the stride=2) halves the input grid's resolution.
Thus, every discriminator has the same receptive field ($11\times 11 \times 11$). The output of the discriminator is
a 3D score map in which every element classifies the corresponding $11\times 11 \times 11$ patch of the input voxel grid to be fake or real.

Unlike the generators, the discriminators can afford to use 3D convolutions on 3D voxel grids,
as they consume much less memory due to following reasons. 
1) With the stride=2 in the first layer, the input grid's resolution is halved.
2) More importantly, the discriminators are only needed at training time, and as we will describe next, they are trained sequentially from
low level to high level. At any moment in time, only one discriminator needs to be kept in memory.
In contrast, all the generators in the GAN hierarchy must be stacked together in order to generate a shape.

\subsection{Training}\label{sec:training}
We train our generators and discriminators in a progressive manner, from the lowest level (the coarsest scale)
to the highest level (the finest scale).
The loss function for training the
$i$-th level includes an adversarial term and a reconstruction term, namely,
\begin{equation}
	\underset{G_i, M_i}{\min}\underset{D_i}{\max}\;\Ladv(G_i, M_i, D_i)+\alpha\Lrec(G_i, M_i),
\end{equation}
where the weighting factor $\alpha$ is set to $10$ in all our experiments.  The
projection module $P$ is also jointly trained in the lowest level ($i=0$)
and kept fixed for the rest of the training.  
Once the training for the $i$-level is finished, the weights of $G_i$ and $M_i$ stay fixed, and $D_i$ is discarded.  
In addition, the weights of $M_i$ and $D_i$ will be used to initialize the training of $M_{i+1}$ and $D_{i+1}$.

For the adversarial term $\Ladv$, 
since the discriminator outputs a score map that classifies each of the $11\times 11 \times 11$ patches of the voxel grid,
we treat the average of the score map as the final discriminator score and use the WGAN-GP
\cite{gulrajani2017improved} training objective, which we found through experiments produces the best results.

The reconstruction term $\Lrec$ requires that a specific sequence of noise $\{Z_0^*, z_1^*, ..., z_N^*\}$ is able to recover the input example at each scale.
In our experiments, we set $\{z_1^*...,z_N^*\}$ to zeros, and $Z_0^*$ to be a fixed random 3D noise (i.e., drawn once and then kept fixed).
The reconstruction loss is defined using the mean squared error,
\begin{equation}
	\Lrec = ||\xrec{i}-x_i||^2,
\end{equation}
where $\xrec{i}$ is the generated shape at scale $i$ when using the specific noise sequence $\{Z_0^*, 0,...,0\}$.

Following suggestion by Shaham et al.~\shortcite{shaham2019singan}, we also use
the reconstructed shape to determine the standard deviation $\sigma_i$ of the
Gaussian noise $z_i$ (one that is injected to the generator in each level).  Specifically, we set 
\begin{equation}
	\sigma_i = \hat\sigma\; ||(\xrec{i-1})\upsample - x_i||,\quad i >0,
\end{equation}
where $\hat\sigma$ is a predefined value and we use $\hat\sigma=0.1$ in our experiments.
For the initial 3D noise $Z_0$ at the coarsest scale, we simply set $\sigma_0=1.0$.

\subsection{Implementation Details}
\label{sec:implementation}
When constructing the voxel pyramid $\{x_0, ..., x_N\}$ of the input shape, we set the downsampling factor $r=0.75$ in all our experiments.
We choose the number of scales $N$ such that the largest dimension of $x_0$ is around $22\sim33$ (\ie, $2\sim3$ times larger than the receptive field of the discriminator).
For example, we set $N$ to be $6$ or $7$ for the resolution $s_N=128$.
If the smallest dimension of any $x_i$ is less than $15$ voxels, we resize that dimension to be $15$. 
In addition, we apply a Gaussian filter with $\sigma=0.5$ to each $x_i$ to smooth its boundary.

Throughout our experiments, we set the number of tri-plane feature channels
$C=32$.  Each MLP decoder $M_i$ has one hidden layer of dimension $32$, and we
apply Sigmoid function to restrict its output within the range $[0,1]$.  Each
convolution block in the generator $G_i$ and discriminator $D_i$ follows the
form of Conv-InstanceNorm-LeakyReLu \cite{ulyanov2016instance}, except the last
one, which has no normalization layer and activation function.  All
convolutional layers use number of channels 32. The architecture details are included in
the \tabref{network} of the appendix.

We train every GAN for $2000$ iterations using the Adam optimizer \cite{kingma2014adam} with a learning rate of $1e{-4}$ and $\beta_1=0.5$.
Each training iteration includes three discriminator updates followed by three generator updates. 
For the WGAN-GP objective, we set its gradient penalty weight to be 0.1.
Our method is implemented using PyTorch \cite{paszke2019pytorch} and trained on an Nvidia GeForce RTX 3090 GPU.
With these settings, it takes about $4$ hours to train our model (at resolution of $256$ along its largest dimension), 
and less than $0.1$ second for generating a shape at inference time.

\section{Evaluation Results}
\begin{figure*}
  \centering
  \includegraphics[width=0.99\textwidth]{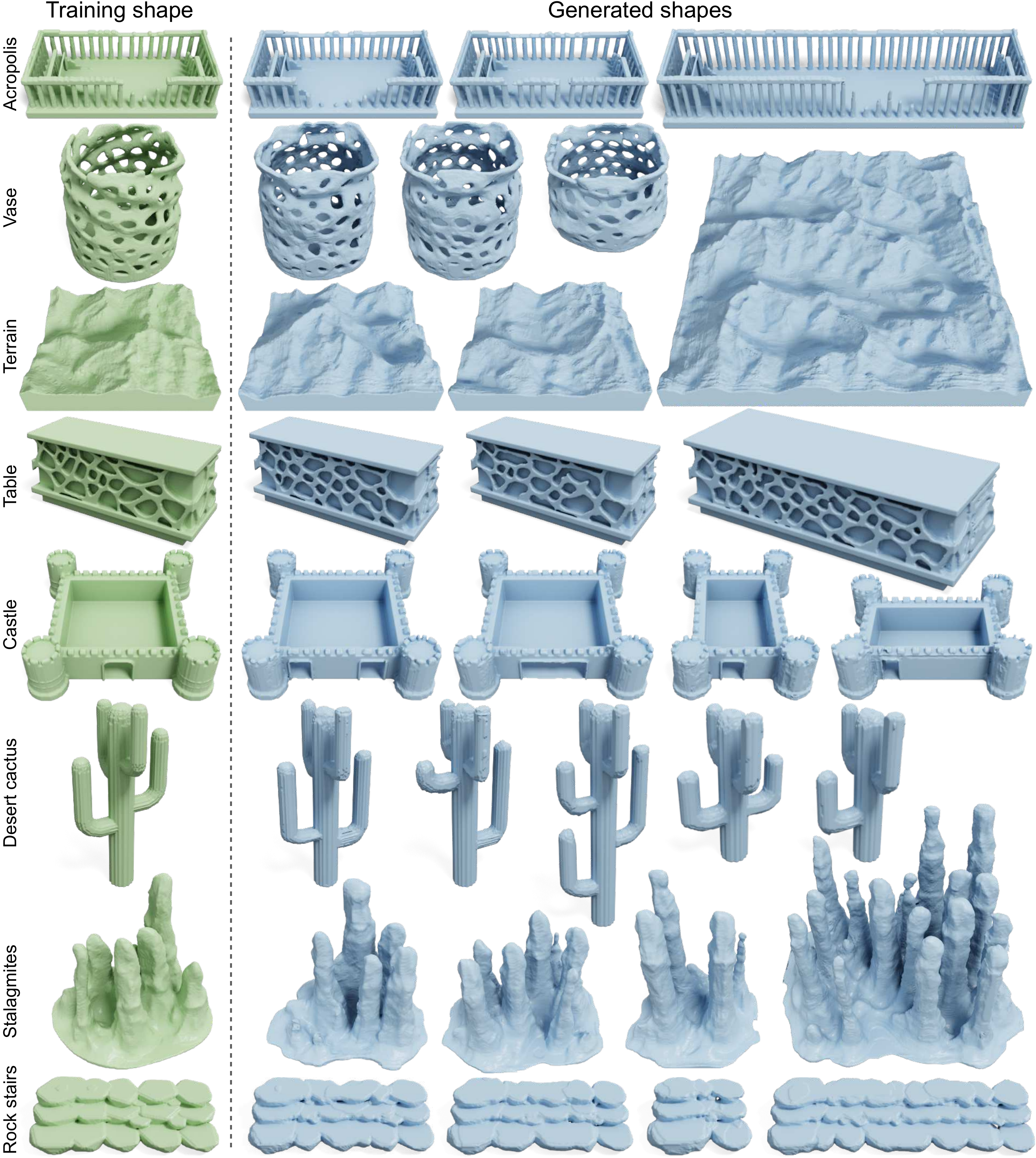}
  \vspace{-1mm}
  \caption{\textbf{Randomly generated shapes}. 
  For each training shape, we show two randomly generated shapes that have the same spatial dimensions and additional shapes
  having different sizes and aspect ratios.
  Here, the largest spatial dimension of each training shape is $256$, except
  for the vase and stalagmites example. For those two shapes, the largest spatial dimension is $192$. 
  From top to bottom, \copyright Acropolis by choly kurd under Standard License (turbosquid.com), \copyright Vase by davidmus under CC BY-SA, \copyright Terrain by BOXX3D under Editorial License (cgtrader.com), \copyright Table by akerStudio under RF, \copyright Castle by BlackMotion under RF, \copyright Desert cactus by exnihilo under RF, \copyright Stalagmites by wernie under RF, and \copyright Rock stairs by Misanthropiclion under RF.
  }
  \label{fig:gallery}
\end{figure*}

We now present our evaluation experiments and results.
The generated 3D shape is represented as a neural implicit function, which
we visualize in a standard process:
we extract the surface mesh using Marching Cubes algorithm~\cite{marchingcubes}
followed by Laplacian Smoothing~\cite{vollmer1999improved} to reduce Marching Cubes' artifacts.
In \tabref{data_config} of the appendix, we list the training resolution and number of scales for all shape eaxmples in the paper.

A gallery of generated shapes is shown in \figref{gallery}, and more results 
are provided in \figref{gallery2} and \figref{gallery3} of the appendix.
We also provide an offline webpage in the supplementary material for interactive view of some example results.  
For each example, we show two generated samples that have the same bounding box size
as the input shape, as well as samples that have different sizes and aspect ratios.
As can be seen, all our generated shapes preserve the global structure of the 
input shape,
while presenting rich local variations.
Remarkably, the structural preservation and local variation
adapt to input bounding box sizes and aspect ratios, allowing the user to quickly generate an ample assortment of 
shapes from a single input.

\subsection{Comparison}
\label{sec:comparison}

\begin{table*}[]
\caption{\textbf{Quantitative evaluation.} $\uparrow$: a higher metric value is better; $\downarrow$: a lower metric value is better. 
	We compare the generation quality of each method using three metrics. 
	The last column is the average score over the ten testing examples.}
\label{tab:comparison}
\secspace
\begin{tabular}{llccccccccccl}
\toprule
\multirow{2}{*}{Metrics}    & \multirow{2}{*}{Methods} & \multicolumn{11}{c}{Examples}                                                                   \\
&                          & Acropolis    	& Terrain 		 & Stalagmites    & Stairs 			& Rock  		 & Wall  		  & Vase  		   & Cheese 		 & Cactus 		  & Tree  		   & Avg. \\
\midrule
\multirow{3}{*}{LP-IoU
			$\uparrow$}     
& Ours                     & \textbf{0.902} & \textbf{0.647} & \textbf{0.254} & \textbf{0.777}  & \textbf{0.004} & \textbf{0.848} & \textbf{0.175} & \textbf{0.564}  & \textbf{0.583} & \textbf{0.015} & \textbf{0.477} \\
& SinGAN-3D                & 0.635     		& 0.645   		 & 0.253          & 0.682  			& \textbf{0.004} & 0.772 		  & 0.173 		   & 0.524  		 & 0.343  		  & 0.001 		   & 0.403 \\
& DGTS                     & 0.015	   		& 0.207	 		 & 0.013	      	  & 0.110			& 0.001			 & 0.155		  & 0.004		   & 0.100	 		 & 0.028  		  &	0.001 		   & 0.063  \\
\midrule
\multirow{3}{*}{LP-F-score
			$\uparrow$} 		
& Ours                     & \textbf{0.950} & 0.812   		 & 0.363          & \textbf{0.865}  & \textbf{0.129} & \textbf{0.892} & \textbf{0.214} & \textbf{0.716}  & \textbf{0.847} & \textbf{0.149} & \textbf{0.594} \\
& SinGAN-3D                & 0.783     	    & \textbf{0.823} & \textbf{0.373} & 0.781  			& 0.125 		 & 0.847 		  & 0.207 		   & 0.673  		 & 0.642  		  & 0.048 		   & 0.530 \\
& DGTS                     & 0.018	   		& 0.382	 		 & 0.111	      	  & 0.197			& 0.101			 & 0.231		  & 0.007		   & 0.300	 		 & 0.243  		  & 0.059 		   & 0.165 \\
\midrule
\multirow{3}{*}{SSFID
			$\downarrow$}   
& Ours                     & \textbf{0.037} & \textbf{0.050} & \textbf{0.078} & 0.102  			& \textbf{0.020} & \textbf{0.272} & \textbf{0.029} & \textbf{0.065}  & \textbf{0.018} & \textbf{0.073} & \textbf{0.074} \\
& SinGAN-3D                & 0.065     		& 0.058   		 & 0.089          & \textbf{0.093}  & 0.035 		 & 0.475 		  & 0.059 		   & 0.145  		 & 0.033  		  & 0.123 		   & 0.118 \\
& DGTS                     & 3.62      		& 2.08	 		 & 1.81	          & 2.00			& 0.151			 & 3.67			  & 1.95		   & 1.36	 		 & 0.403  		  & 0.201 		   & 1.724 \\
\bottomrule
\end{tabular}
\secspace
\end{table*}

\begin{figure*}
  \centering
  \includegraphics[width=\textwidth]{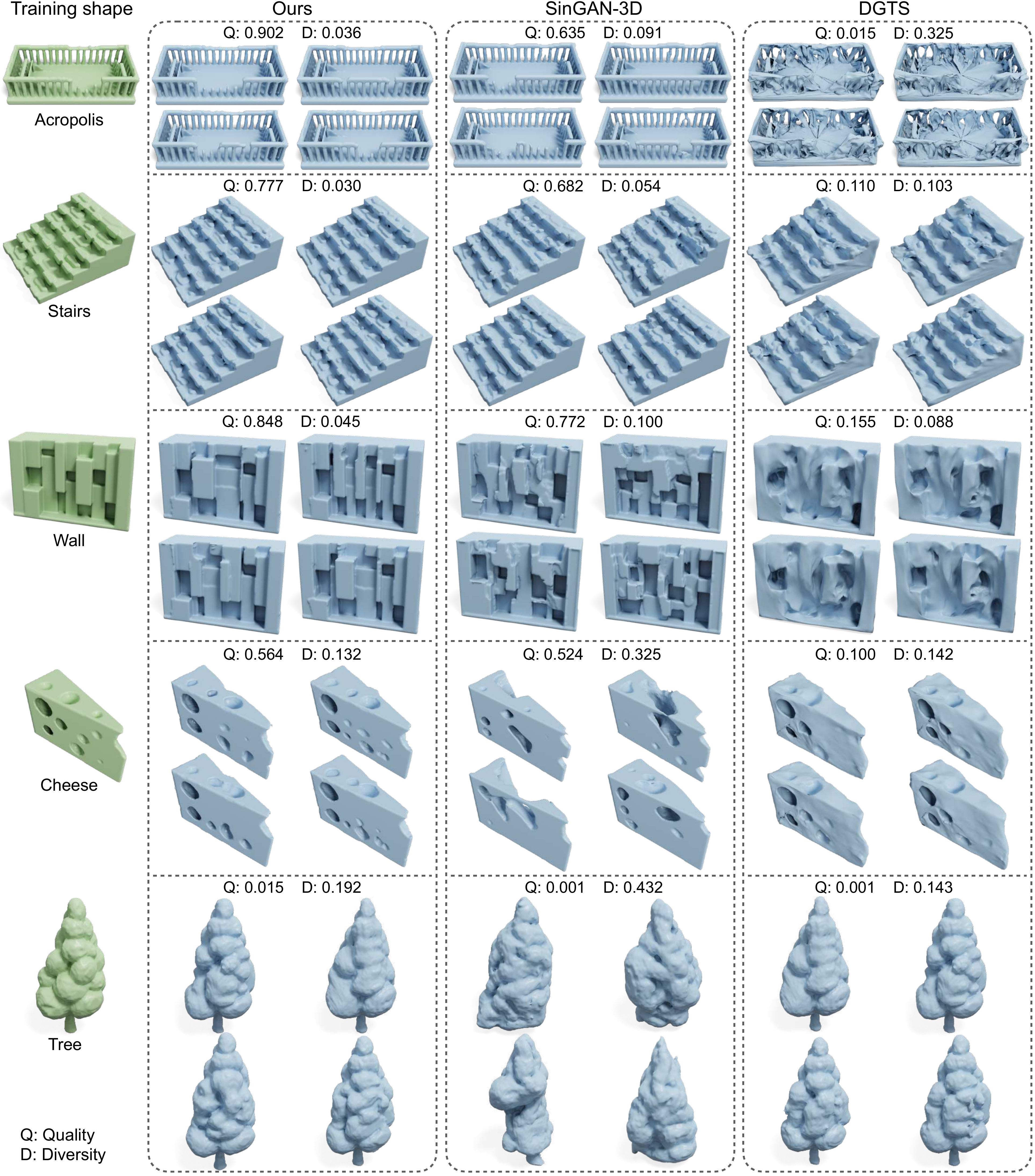}
  \vspace{-8mm}
  \caption{\textbf{Visual comparison on our testing examples.} 
  	Due to the limited space, here we only show comparison results on five testing examples and include the rest in \figref{full_comp2} of the appendix.
  	The quality score (Q) is LP-IoU defined in ~\secref{comparison}.
  	The diversity score (D) is defined as the pairwise difference ($1 - \text{IoU}$) within a set of generated shapes (see \appref{supp_metric} for detailed definition). 
  	For each method, we randomly generate $100$ shapes to calculate both scores.
  	From top to bottom, \copyright Acropolis by choly kurd under Standard License (turbosquid.com), \copyright Stone stairs by ivanzubak under Standard License (turbosquid.com), \copyright Wall by stray under RF,\copyright Cheese by Thiennguyen2106 under RF, and \copyright Tree by Ada\_King under Standard License (turbosquid.com).
  	}
  \label{fig:comparison}
\end{figure*}

\paragraph{Compared methods}
We compare our method against two baselines.
\textbf{SinGAN-3D} is an approach that we attempted by changing the 2D convolutions in 
SinGAN~\cite{shaham2019singan} into 3D convolutions (to support 3D shape generation).
For fair comparison, our implementation of SinGAN-3D uses the same discriminator structure and training strategy as used in our method.
The only difference is that SinGAN-3D uses a 3D convolutional generator 
{operating directly on voxel grids}.
\textbf{DGTS} \cite{Hertz2020deep} is a prior 3D generative model working with a single example,
although it aims to learn local geometric textures on mesh surfaces but not global structures.
We use their published source code for comparison.

When comparing our method with SinGAN-3D, 
we set the largest dimension of the training shape to $128$ voxels in both methods, 
and use $6$ or $7$ training scales depending on its aspect ratio.
We note that SinGAN-3D has a much larger memory footprint than our method.
Thus, in our experiments, we can not use it at a resolution as high as what we
use in~\figref{gallery}, because SinGAN-3D causes the GPU to run out of memory.

Also, unlike our method, DGTS takes as input a triangle mesh.
Thus, for fair comparison, 
we use Marching Cubes to create a target mesh from a voxel grid that is also provided to our method.
We then simplify it into a template mesh with
$100$ faces using an edge collapsing algorithm~\cite{garland1997surface}.  With
the target and template meshes, we use their optimization scheme to create the
training meshes in $5$ scales and finally train their model.

\paragraph{Evaluation Metrics}
To measure the quality of generated shapes, we adopt the following metrics.
\textbf{LP-IoU} and \textbf{LP-F-score}~\cite{chen2021decor}, according to their paper, measure the local
{plausibility} of the generated shape.  They are defined as the percentage of
local patches (i.e., the $11\times11\times11$ voxels in our setup) of a generated shape 
that are ``similar'' to at least one patch of the training shape.  
We consider two patches sufficiently ``similar'' if their IoU or F-score is above $0.95$.  
To avoid sampling featureless patches\textemdash patches that are far away from the shape surface\textemdash
we only consider voxel patches across the surface (i.e., having at least one occupied voxel and one unoccupied voxel {in the central $3\times 3\times 3$ area of the patch}).

In addition, \textbf{SSFID} (Single Shape Fr\'echet Inception Distance) 
measures to what extent the generative model captures the patch statistics of the
training shape.  Similar to SIFID (Single Image Fr\'echet Inception
Distance)~\cite{shaham2019singan}, we use the deep features output by the
second convolutional block in a pretrained 3D shape classifier (which we  
take from~\cite{chen2021decor}).  SSFID is defined as the Fr\'echet Inception
Distance (FID) of those deep features between the generated
and example shapes.  Details of the metric evaluation are included in \appref{supp_metric}.

We select $10$ shapes in different categories as testing examples. 
These shapes have different topologies and patch variations across a range of scales.
For each evaluated method and each testing example, we randomly generate $100$ shapes 
and report the average scores of the metrics. For DGTS, we
voxelize their generated meshes for the metric evaluation.

\paragraph{Results}
The evaluation results are reported in ~\tabref{comparison}, and the corresponding shapes are shown in~\figref{comparison} and \figref{full_comp2} of the appendix.
For nearly all testing examples, our method produces the best scores under all three metrics. 
Compared to the results of SinGAN-3D, our generated shapes better preserve the
structure of the input shape (\eg, see the wall example in~\figref{comparison}).
In these tests, DGTS performs the worst due to its inability to 
learn large structures and anisotropic geometric features,
which is indeed one of the limitations acknowledged in their paper~\cite{Hertz2020deep}.
Also, it can not learn well complex topologies (\eg, see the acropolis example in~\figref{comparison}).
In those cases, 
DGTS takes a long time (>$15$ hours) for training meshes preparation and the training process,
possibly due to the large
number of vertices and many self-intersections resulted from their mesh optmization process.

Besides the shape quality, the generation diversity is also of
interest. However, measuring diversity solely without taking into
account quality makes little sense.
As an extreme example, a set of shapes with strong random noise is probably considered diverse (or different from each other), 
but those shapes are useless because of the poor quality.
As an attempt, we measure diversity using the pairwise difference within a set of random generated shapes
shown in \figref{comparison}.
Under this metric, SinGAN-3D usually obtains a better score,
although its output shapes have lower quality than ours. 

Another advantage of our method over SinGAN-3D is its efficiency, 
because unlike SinGAN-3D, our generator requires no 3D convolution.
Here, we use both SinGAN-3D and our model to generate shapes and examine their subsequent GPU memory and time costs.
When executing our model, we query all grid points through the MLP network in a single forward pass.
The results are presented in \tabref{cost}, which shows that
our model requires much less GPU memory and runs faster than SinGAN-3D as the output resolution increases.
We can further reduce the GPU memory footprint by using algorithms such as Multi-resolution IsoSurface Extraction (MISE)~\cite{Mescheder_2019_CVPR}.

\begin{table}[t]
\caption{\textbf{Comparison of GPU memory and time cost.}
        OOM: out of memory. Since both methods are multi-scale generative models, 
        in this experiment,
        we use $6$, $7$ and $8$ training scales 
        for output resolutions $128^3$, $192^3$ and $256^3$, respectively. 
        Inference time is the average over $100$ trials.
}\label{tab:cost}
\secspace
\begin{tabular}{lccc}
\toprule
Methods                    & \begin{tabular}[c]{@{}c@{}}Output\\ Resolution\end{tabular} & \begin{tabular}[c]{@{}c@{}}GPU\\ Memory\end{tabular} & \begin{tabular}[c]{@{}c@{}}Inference\\ Time\end{tabular} \\ \hline
\multirow{3}{*}{Ours}      & $128^3$ & 4.06G  & 0.041s \\ 
                           & $192^3$ & 7.27G  & 0.051s \\ 
                           & $256^3$ & 19.34G & 0.057s \\
\midrule
\multirow{3}{*}{SinGAN-3D} & $128^3$ & 9.31G  & 0.082s \\ 
                           & $192^3$ & 20.45G & 0.453s \\ 
                           & $256^3$ & OOM    & N/A      \\
\bottomrule
\end{tabular}
\end{table}

\subsection{Ablation Study}
\label{sec:ablation}

\begin{table}[h]
\caption{\textbf{Ablation study}. Metric values shown here are averages over the $10$ testing examples, under the same setting as ~\tabref{comparison}.
	Metric values for each testing example are available in \tabref{full_ablation} of the appendix.
}\label{tab:ablation}
\secspace
\begin{tabular}{lccc}
\toprule
                     & LP-IoU $\uparrow$ & LP-F-score $\uparrow$ & SSFID $\downarrow$ \\
\midrule
Proposed Method      & 0.479  & 0.594      & 0.072 \\
Initial 2D Noises    & 0.438  & 0.554      & 0.086 \\
No Gaussian Blur     & 0.456  & 0.564      & 0.084 \\
No Weights Reuse     & 0.451  & 0.566      & 0.078 \\
\bottomrule
\end{tabular}
\end{table}

We conduct ablation studies to validate the design choices of our model.  In
\tabref{ablation}, we compare our proposed method with the following variants:

\emph{Initial 2D noises}, in which we independently sample three 2D noise maps
to construct the input tri-plane map to $G_0$. In this variant, the projection network is not needed any more. 
In contrast, our method projects a sampled 3D noise onto the three feature planes.

\emph{No Gaussian Blur}, in which we do not apply Gaussian filter to blur the
voxel grid of the input shape (recall \secref{implementation}).  
Without the Gaussian blur, the grid values would be either $0$ or $1$, indicating 
whether or not each grid point is occupied by the shape. 
As a result, the discriminator can trivially distinguish a synthesized shape, because the grid data
decoded by the MLP network will likely have values between 0 and 1\textemdash a clue for the discriminator
to do its job without carefully examining the shape features.
Such a discriminator, although effective for distinguishing synthesized shapes, offers little help
for improving the generator.

\emph{No Weights Reuse}, in which we randomly initialize the weights of the GAN
in each level at the beginning of the training process.
On the contrary, our proposed method initializes the
network weights of a particular level using the weights of a lower level GAN (recall \secref{training}). 

Our method outperforms all these variants (see \tabref{ablation}), 
and thus our design choices are well justified.

\subsection{Choice of Voxel Pyramid Resolution}\label{sec:choice_res}
\begin{figure}[t]
  \centering
  \includegraphics[width=0.99\linewidth]{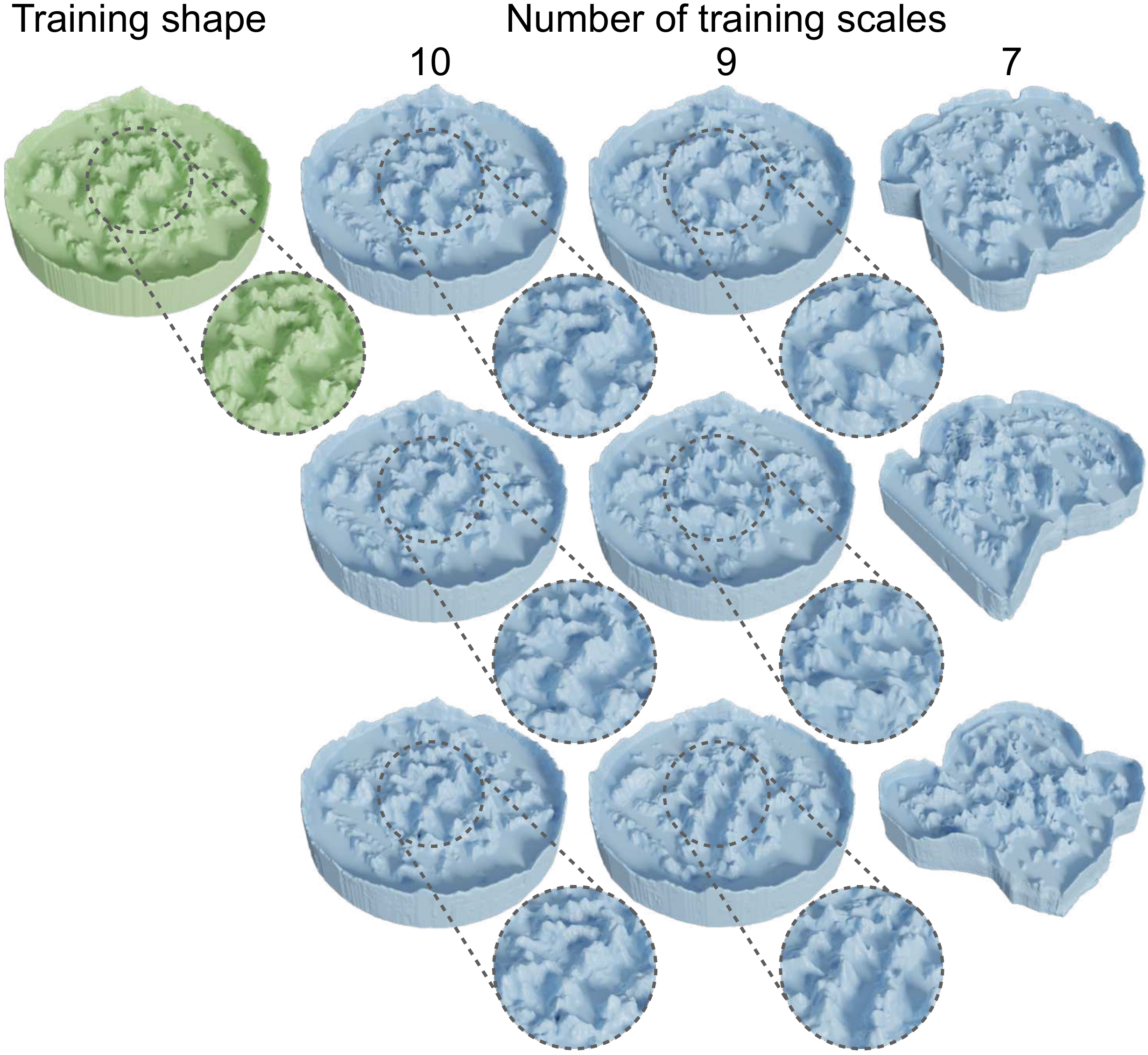}
  \vspace{-1mm}
  \caption{\textbf{Choice of the coarsest scale.} In this example, we fix the finest level resolution ($256 \times 54 \times 256$) 
  and the downsampling factor $r$ is 0.75. 
  We build three voxel pyramids of the input shape using different numbers of scales, and thus their coarsest level resolution varies.
  From left to right, the numbers of scales are 10, 9, and 7, respectively, and the 
  coarsest level resolutions are $20\times 5\times 20$, $27\times 6\times 27$ and $46\times
  10\times 46$.  
  In each setting, we show three random generated samples from the trained generative models.
   We also show the shapes' local variations in the inset figures.
   See discussion in \secref{choice_res}.
  \copyright Mountain landscape with water by vis-all-3d under Editorial License (cgtrader.com).
  }
  \label{fig:training_scales}
\end{figure}
Recall that we build a voxel pyramid of the input example to train our generative model.
Here we conduct an experiment to understand how the range of scales covered by
the pyramid affects the generation results.

The experiment is shown in \figref{training_scales}, where the finest level
resolution of the voxel pyramid stays fixed, but the number of scales varies.
When the number of scales is too large, the coarsest level resolution is low,
and every voxel in the coarsest level grid occupies a large spatial region.  As
a result, the discriminator at this level (whose receptive field is always
$11\times11\times11$) examines the shape features in a large region, too large to allow any structural 
variation (see the second column of \figref{training_scales}).
The generated shapes simply overfit the example shape.
On the other extreme, when the number of scales is too small, 
the coarsest level resolution is still high, and thus
a voxel in the coarsest level grid occupies only a small region. Consequently, the discriminator is
unable to force the generator to learn large-scale structures, and the
generated shape appear to be structurally incoherent (see the fourth column of
\figref{training_scales}).

As a rule of thumb, in practice, we typically choose the number of scales
such that the largest dimension of the coarsest grid is about two or three
times larger than the discriminator's receptive field.

\subsection{Shape Interpolation and Extrapolation}
\label{sec:interpolation}
\begin{figure}[t]
  \centering
  \includegraphics[width=\linewidth]{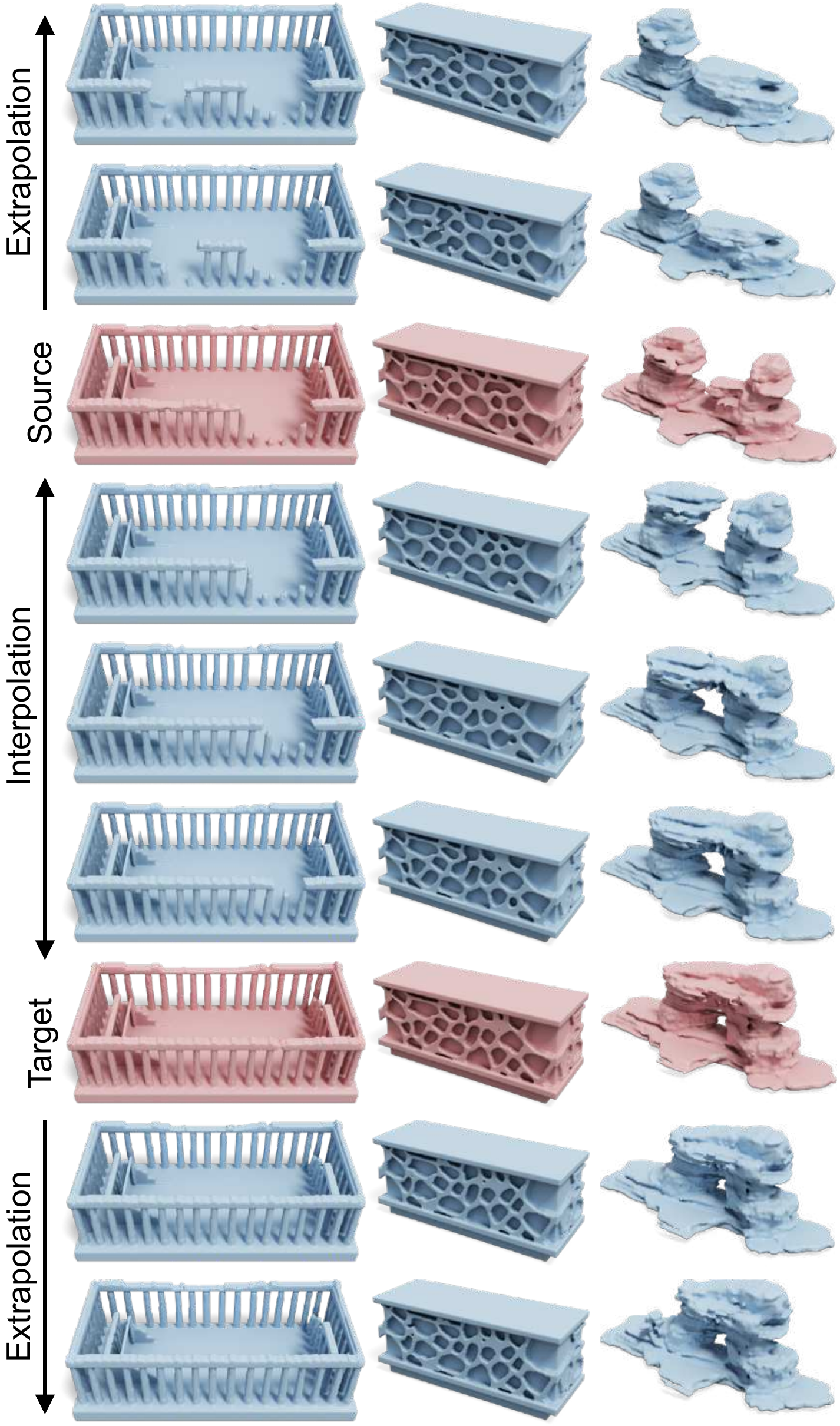}
  \caption{\textbf{Shape interpolation and extrapolation.} 
 We perform shape interpolation and extrapolation by linearly blending the input noises of the source and target shapes.
 From top to bottom, the blending weights are $-0.5, -0.25, 0, 0.25, 0.5, 0.75, 1, 1.25$, and $1.5$, respectively.
 Note the smooth transition across the source and target shapes.}
  \label{fig:interpolation}
\end{figure}

Our generative model naturally supports shape interpolation and to a certain extent extrapolation.
Consider two shapes generated from two initial 3D noises $Z_0^1$ and
$Z_0^2$, respectively. The shape interpolation is straightforward, \ie, by feeding into the generative model a linearly
interpolated noise $Z_0^\alpha=(1-\alpha)\cdot Z_0^1+\alpha \cdot Z_0^2$, 
where $\alpha$ is the interpolation parameter.
In fact, $\alpha$ is not necessarily in the range of $[0, 1]$;
we can extend it to be negative or greater than one for shape extrapolation.
Here, noise at other scales ($z_1, ..., z_N$) stay fixed.

We show three examples in~\figref{interpolation}, and include more results in
\figref{interpolation2} of the appendix.  
All the results present smooth transition across two generated shapes provided as input.
Notably, in the acropolis example in \figref{interpolation}, the breakage of the front columns
is gradually closed and filled as the interpolation changes from the source to the target shape.
This is not possible via simple interpolation in voxel space.  
In addition, %
the extrapolation towards the source direction produces more breakage, 
while extrapolating towards the target direction retains the completeness of the columns.
These results suggest that our generative model is able to learn a smooth and meaningful mapping from random noise to realistic shapes.

\subsection{Higher Resolution Output}\label{sec:highres}
\begin{figure}[t]
  \centering
  \includegraphics[width=\linewidth]{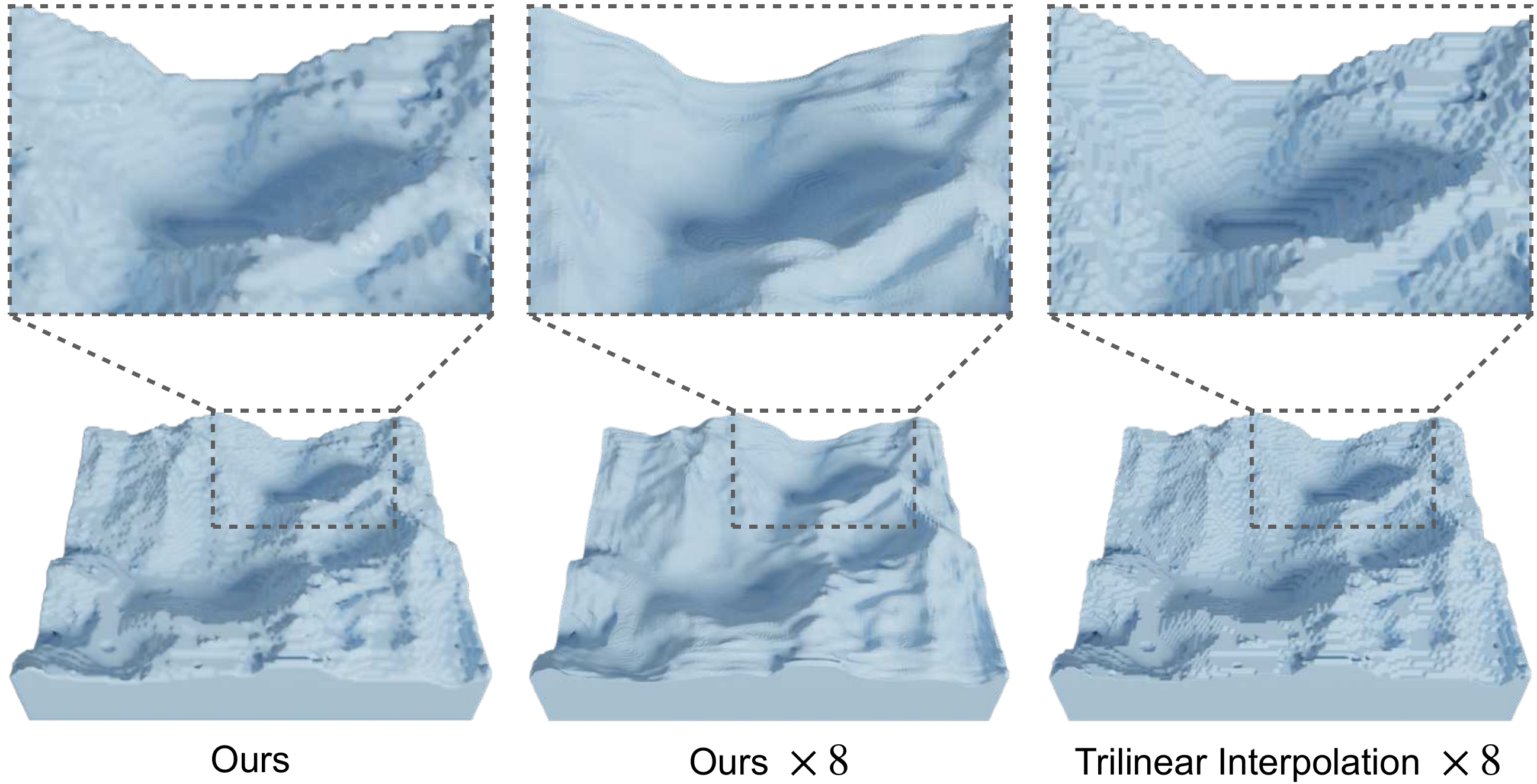}
  \caption{\textbf{Querying at a higher resolution.} Left: A generated sample at
  training resolution ($112\times36\times128$). Middle: $\times 8$ upsampling
  by querying $o(p)$ (resulted from the generated tri-plane representation)
  on a voxel grid with a $\times 8$ resolution.
  Right: $\times 8$ upsampling of the output voxel grid via
  trilinear interpolation. In this figure, visualized meshes are the direct output
  from Marching Cubes. More examples are included in \figref{highres2} of the appendix.}
  \label{fig:highres}
\end{figure}
The neural implicit function, enabled by the MLP network $M_N$ in our model,
allows the model to discretize the output shape at an arbitrary resolution.
We demonstrate this by showing a $\times 8$ upsampling result in~\figref{highres}.  
Here, the output shape is discretized by a grid whose
resolution is 8 times higher than the input grid's resolution.  We compare
the result with a simple upsampling strategy based on trilinear interpolation.
The experiment shows that our result is much smoother while retaining sharp features. 
However, the ability to query at an arbitrary resolution does not
imply arbitrary geometric details. 
The level of details that our model can synthesize is still limited by the training resolution.

\subsection{Failure Cases}
\label{sec:limitation}
\begin{figure}[t]
  \centering
  \includegraphics[width=0.98\linewidth]{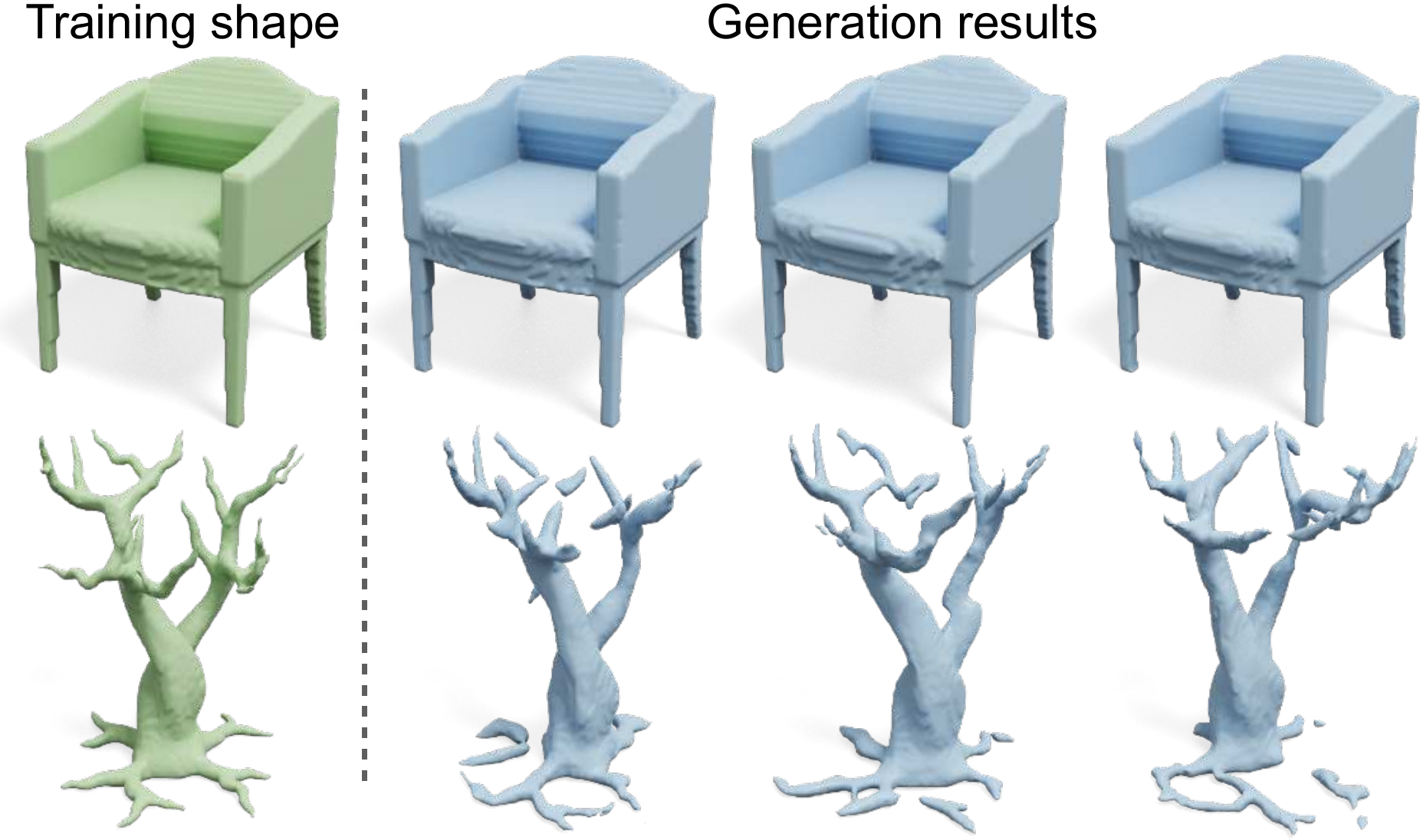}
  \vspace{-1mm}
  \caption{\textbf{Failure cases.} Top: Our method overfits to the training
  shape that lacks rich local patterns. Bottom: Like other voxel-based generative model, 
  it can not learn sparse thin structures well.
  \copyright Tree branches by 10DollarsAnimation under RF.
  }
  \label{fig:failure}
\end{figure}

Although our generative model can learn from many different types of shapes, 
certain shapes still pose challenges to our model.
1) If the example shape does not have rich local features\textemdash such as a simple chair, which presents
predominantly an overall structure (\figref{failure} top)\textemdash
then our model is unable to learn and generate local variations, and it simply overfits the large structure.
2) Voxel-based deep generative models typically struggle to learn and generate
thin structures, and ours is no different (\figref{failure} bottom).
Primitive guidance~\cite{tang2019skeleton} may help the generative model to learn thin structures. We leave it for future work.

\begin{figure}[t]
  \centering
  \includegraphics[width=0.99\linewidth]{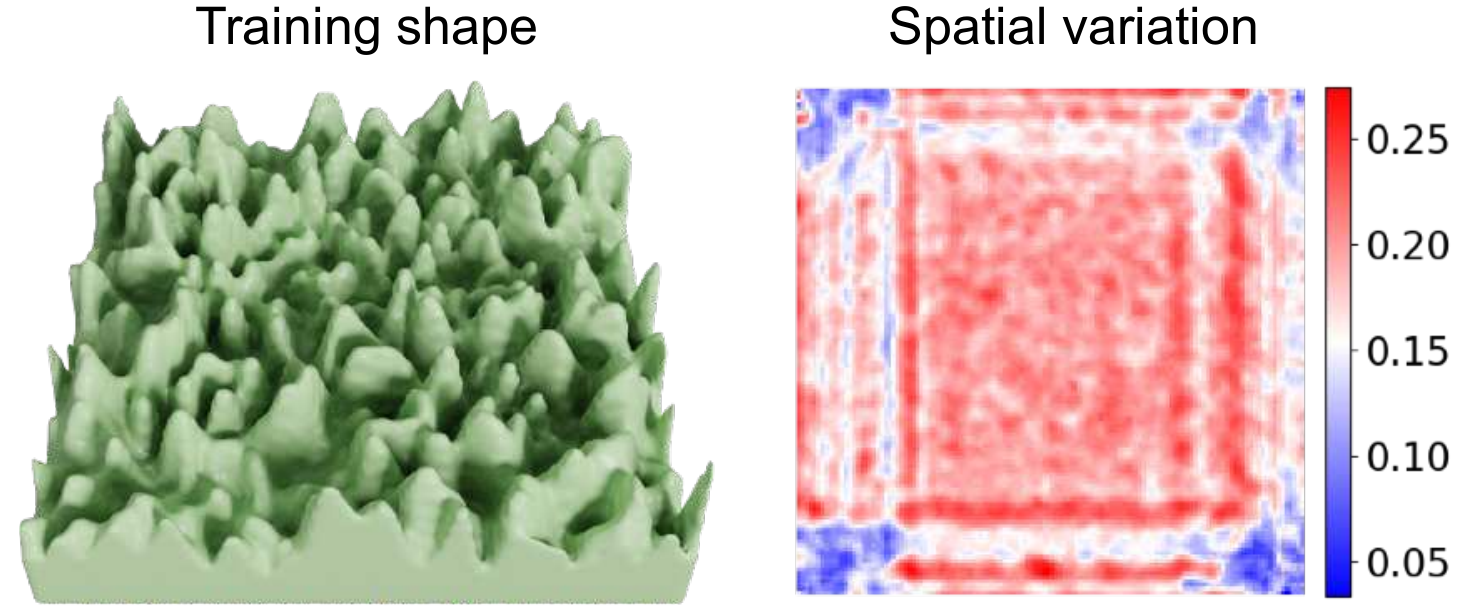}
  \vspace{-2mm}
  \caption{\textbf{Spatial bias.} 
   To reveal spatial bias of our method, we use a noise-like height field as the example 
   shape (left).
   We generate 100 new shapes using our model and compute a spatial variation map (right).
   To this end,
   we compute voxel-wise standard deviations over the generated shapes, and average the values over the $z$-axis
   to obtain the colormap.
   We note much less variation near corners, even though the
   training shape has nearly uniformly distributed geometric features.  
   \copyright Strange terrain by aaron\_g\_randall under Personal Use License (free3d.com).
  }\label{fig:bias}
  \vspace{-1mm}
\end{figure}

\section{Limitations and Future Work}
\label{sec:discussion}
In this work, we propose a generative model that learns from a single example. 
Our method contrasts starkly to existing 3D shape generative models that require a large
set of class-specific 3D shapes for training.  
Trained on a voxel pyramid of the input shape and operating on a hierarchy of tri-plane feature maps, 
our model is able to produce diverse shape variations across different bounding box sizes and aspect ratios.
Meanwhile, all the generated shapes share similar large-scale structures as the input shape.

Our method also has some limitations.
As discussed in recent works~\cite{xu2021positional,choi2021toward}, zero
boundary padding of the convolutional layers serves as an implicit spatial bias. 
Leveraging this bias, a convolutional generator can learn to produce
globally-structured samples from spatially i.i.d.~noises.  Meanwhile, this bias
leads to unbalanced output variation. For example, the output samples have less
variation near the corners, because the zero-padding pattern is unique in
corner regions. We reveal this limitation in our method through a purposely constructed example (see \figref{bias}).
We also tested the sinusoidal positional encoding strategy proposed
in~\cite{xu2021positional}, but find it no better, possibly
because the periodic prior does not apply for 3D shapes.

The tri-plane representation helps reduce the memory and time cost. %
But the highest resolution of the voxel pyramid is still limited (typically $\le 256$),
because the discriminators still perform 3D convolutions on voxel grids.
In future, we may further reduce our model's memory cost by either using 
a more sophisticated training strategy or leveraging a more efficient voxel  
representation (\eg, octree~\cite{Wang-2022-dualocnn}, deformable
tetrahedron~\cite{gao2020learning}).

Another interesting direction is to enable user control in the generation process,
possibly via a conditional input to the network.  For example, 
it would be useful to enable the user to use sketch or brush as a guidance to delete, add, or modify certain
parts of the original shape~\cite{li2020sketch2cad}, while the
generative model automatically makes coherent changes.  
Lastly, we would like to combine the internal (\ie, within the
example) and external (\ie, from other shapes) patch
information~\cite{park2020swapping} for more expressive shape generation.

\begin{acks}
We thank the anonymous reviewers for their constructive feedback. This work was
partially supported by the National Science Foundation (1910839).
\end{acks}

\bibliographystyle{ACM-Reference-Format}
\bibliography{refbib}

\clearpage
\appendix

\section{Network Architectures}

\tabref{network} describes the architecture for each module of our model for a single scale,
and all scales share the same architecture. We will release all the code when the paper is published.

\begin{table}[h]
\resizebox{\columnwidth}{!}{
\begin{tabular}{llccc}
\toprule
Module    					   & Layers                & \makecell{Out \\channels} & \makecell{Kernel \\ size} & Stride \\
\midrule
\multirow{2}{*}{\makecell[l]{Projection 
		\\(for each plane)}}   & \texttt{AdaptAvgPool3D}           & 8 						   & -                         & -         \\
                               & \texttt{Conv2D}                 & 32 						   & $(1, 1)$                  & $(1, 1)$         \\
\midrule
\multirow{4}{*}{\makecell[l]{Generator 
		\\(for each plane)}}   & \texttt{Conv2D+IN+LReLU}       & 32 						   & $(3, 3)$                  & $(1, 1)$         \\
                               & \texttt{Conv2D+IN+LReLU}       & 32 						   & $(3, 3)$                  & $(1, 1)$         \\
                               & \texttt{Conv2D+IN+LReLU}       & 32 						   & $(3, 3)$                  & $(1, 1)$         \\
                               & \texttt{Conv2D}                & 32 						   & $(3, 3)$                  & $(1, 1)$         \\
\midrule
\multirow{2}{*}{MLP Decoder}   & \texttt{Linear+ReLu}           & 32 						   & -                         & -         \\
                               & \texttt{Linear+Sigmoid}        & 1 						   & -                         & -         \\
\midrule
\multirow{3}{*}{Disciminator}  & \texttt{Conv3D+IN+LReLU}       & 32 						   & $(3, 3, 3)$                  & $(2, 2, 2)$         \\
                               & \texttt{Conv3D+IN+LReLU}       & 32 						   & $(3, 3, 3)$                  & $(1, 1, 1)$         \\
                               & \texttt{Conv3D}                & 32 						   & $(3, 3, 3)$                  & $(1, 1, 1)$         \\
\bottomrule
\end{tabular}
}
\caption{\textbf{Network architectures.} 
	\texttt{AdaptAvgPool3D}: adaptive average pooling, whose out channels denote the output size of $x,y$ or $z$-axis.  
	\texttt{IN}: instance normalization layer. \texttt{LReLU}: leaky ReLU with negative slope $0.2$.
}
\label{tab:network}
\end{table}

\vspace{-10mm}
\section{Data Configuration}
In \tabref{data_config}, we list the configuration (\ie training resolution, number of scales) for all training shapes used in the paper.

\section{More Results}
\begin{figure*}
  \centering
  \includegraphics[width=\textwidth]{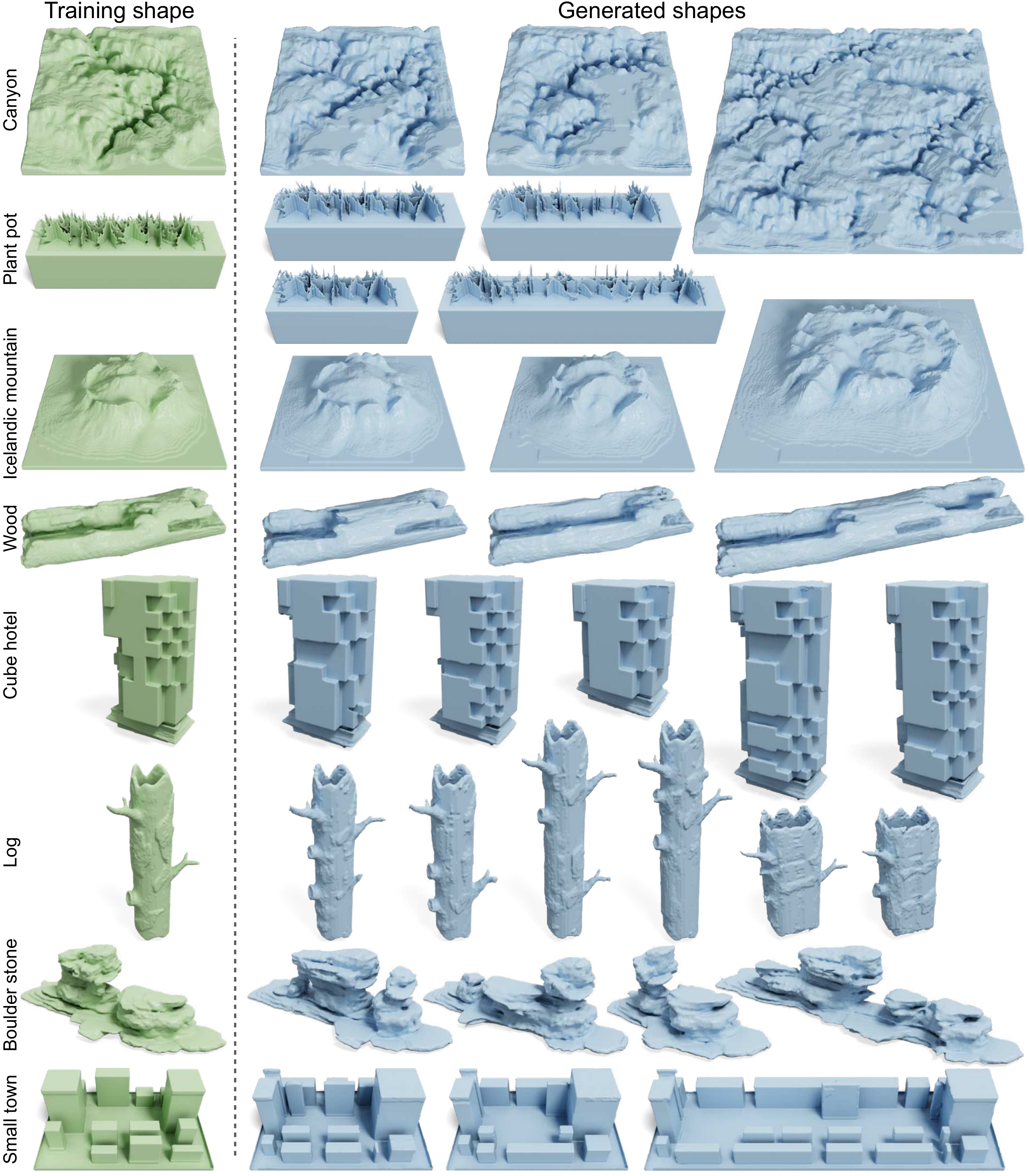}
  \vspace{-8mm}
  \caption{\textbf{More random generation results (1)}. 
  For each training shape, we show two generated shapes of the same spatial dimensions, with the rest having different sizes and aspect ratios.
  From top to bottom, \copyright Canyon by splod67 under CC BY, plant pot from ShapeNet, \copyright Icelandic mountain by saz88 under RF, \copyright Floating wood under RF, \copyright Cube hotel by fredolegros91 under RF, \copyright Log by towercg under RF, \copyright Boulder stone by lml46 under RF, and \copyright Small town by pedram-ashoori under RF.
  }
  \label{fig:gallery2}
\end{figure*}

\begin{figure*}
  \centering
  \includegraphics[width=\textwidth]{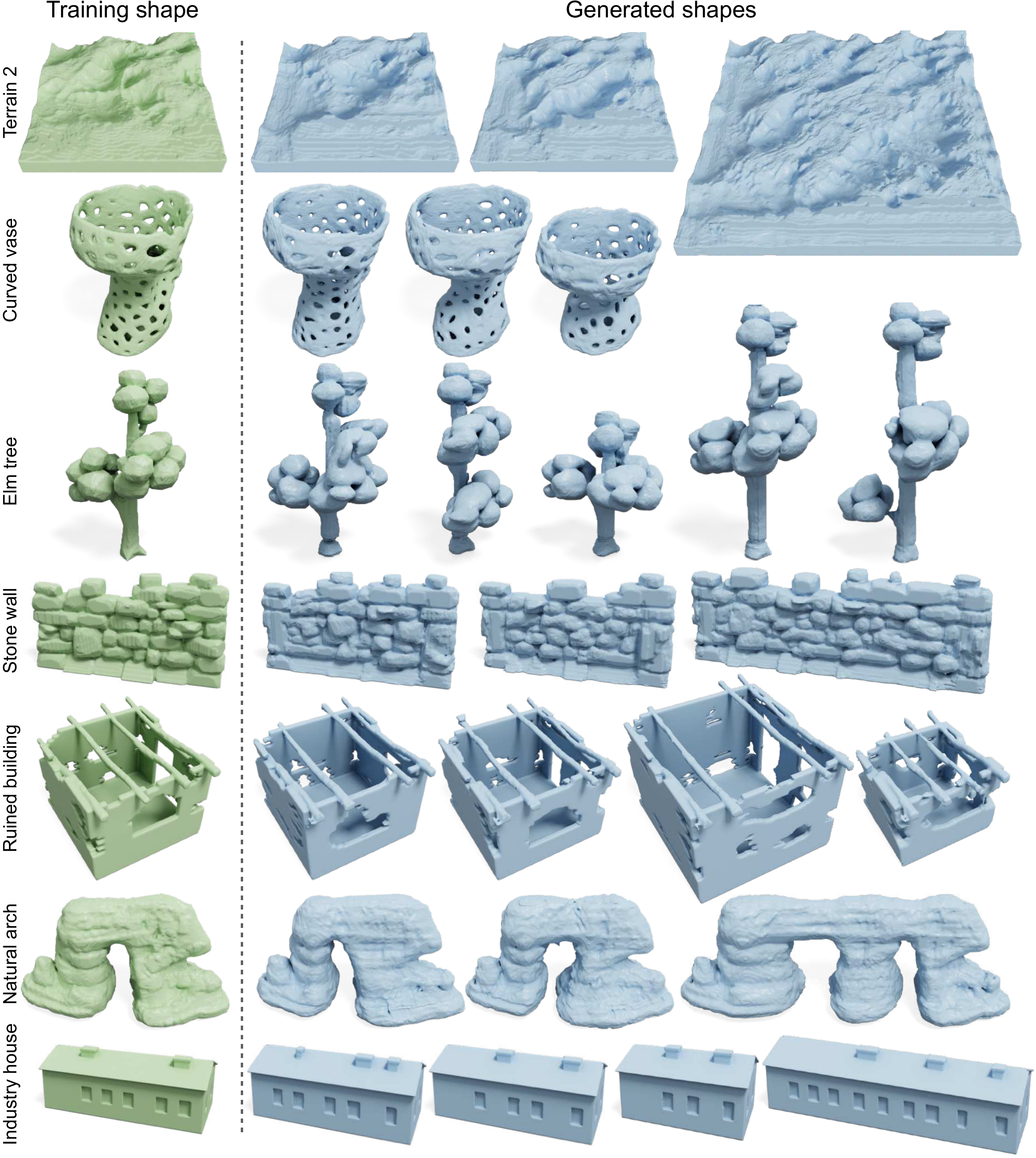}
  \vspace{-8mm}
  \caption{\textbf{More random generation results (2)}. 
  For each training shape, we show two generated shapes of the same spatial dimensions, with the rest having different sizes and aspect ratios.
  From top to bottom, \copyright Terrain 2 by 3dDigital under Standard License (turbosquid.com), \copyright Curved vase by davidmus under CC BY-SA, \copyright Elm tree by darkqueencpn under RF, \copyright Stone wall by bumblebrush under Editorial License (cgtrader.com), \copyright Ruined building by bizkit001 under RF, \copyright Natural arch by smanor under RF, and \copyright Industry house by lukass12 under RF.
  }
  \label{fig:gallery3}
\end{figure*}

\begin{figure*}
  \centering
  \includegraphics[width=\textwidth]{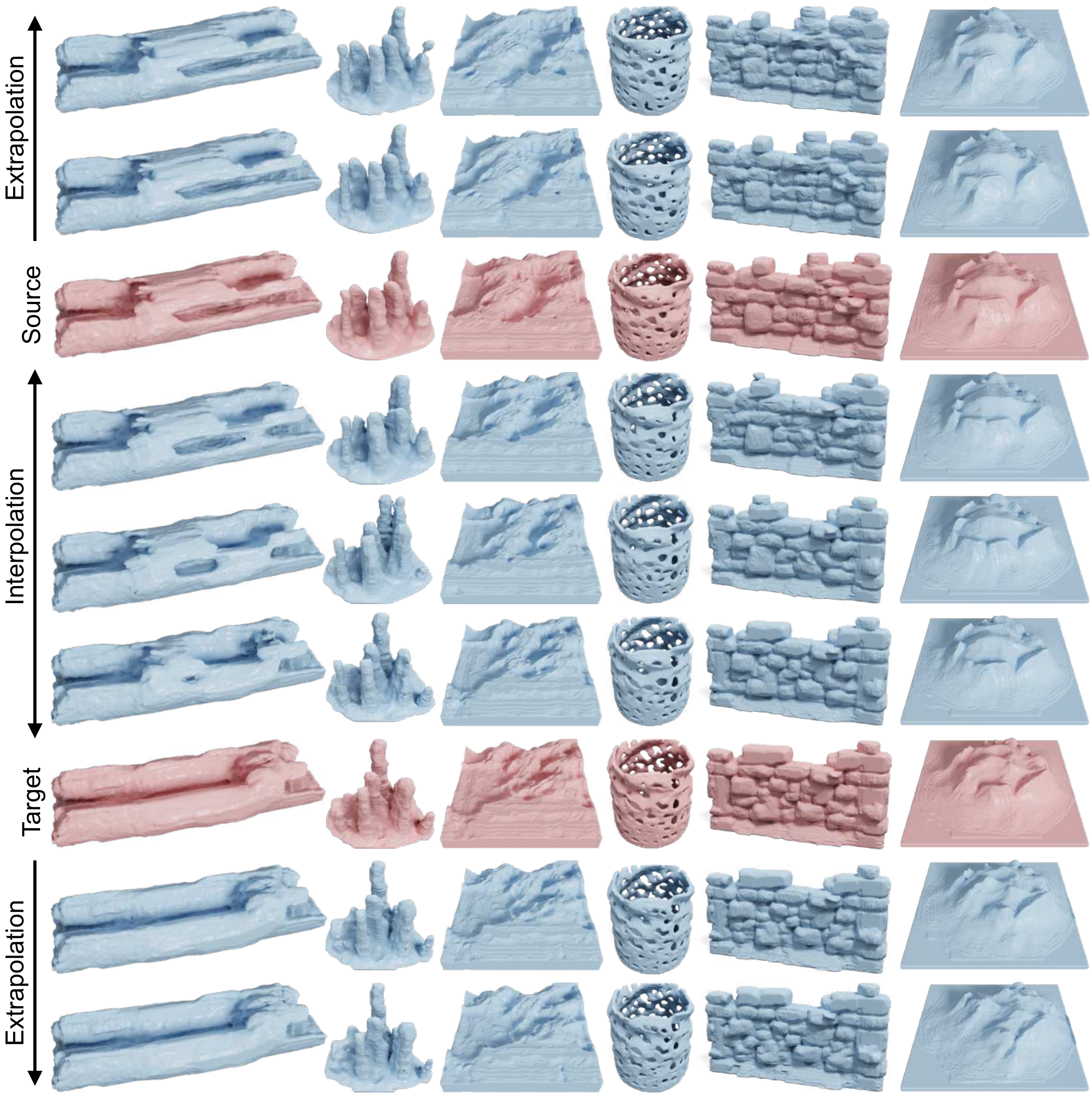}
  \caption{\textbf{More results for shape interpolation and extrapolation.} We perform shape interpolation and extrapolation by linearly blending the input noises of the source and target shapes.
    From top to bottom, the blending weights are $-0.5, -0.25, 0, 0.25, 0.5, 0.75, 1, 1.25$ and $1.5$, respectively.
  	Note the smooth transition across the source and target shapes.}
  \label{fig:interpolation2}
\end{figure*}

\begin{figure*}
  \centering
  \includegraphics[width=\textwidth]{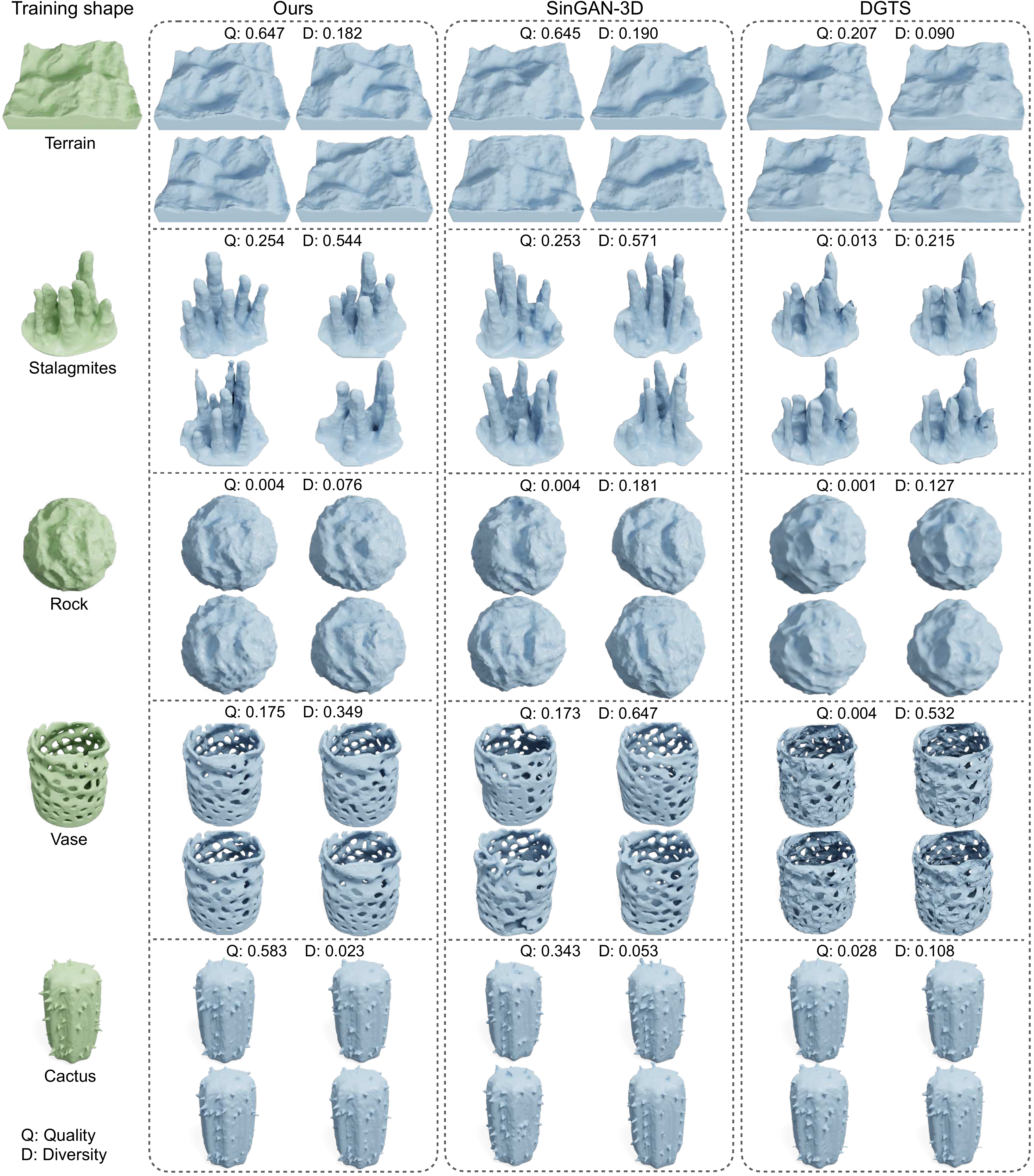}
  \vspace{-8mm}
  \caption{\textbf{Visual comparison on our testing examples.} 
  	The quality score (Q) is LP-IoU defined in ~\secref{comparison}.
  	The diversity score (D) is defined as the pairwise difference ($1 - \text{IoU}$) within a set of generated shapes (see \appref{supp_metric} for detailed definition). 
  	For each method, we generate $100$ shapes to calculate both scores.
  	From top to bottom, \copyright Terrain by BOXX3D under Editorial License (cgtrader.com), \copyright Stalagmites by wernie under RF, \copyright Rock by georgij-space under RF, \copyright Vase by davidmus under CC BY-SA, and \copyright Cactus by m4rios under Standard License (turbosquid.com).
  }
  \label{fig:full_comp2}
\end{figure*}

\begin{figure*}
  \centering
  \includegraphics[width=\textwidth]{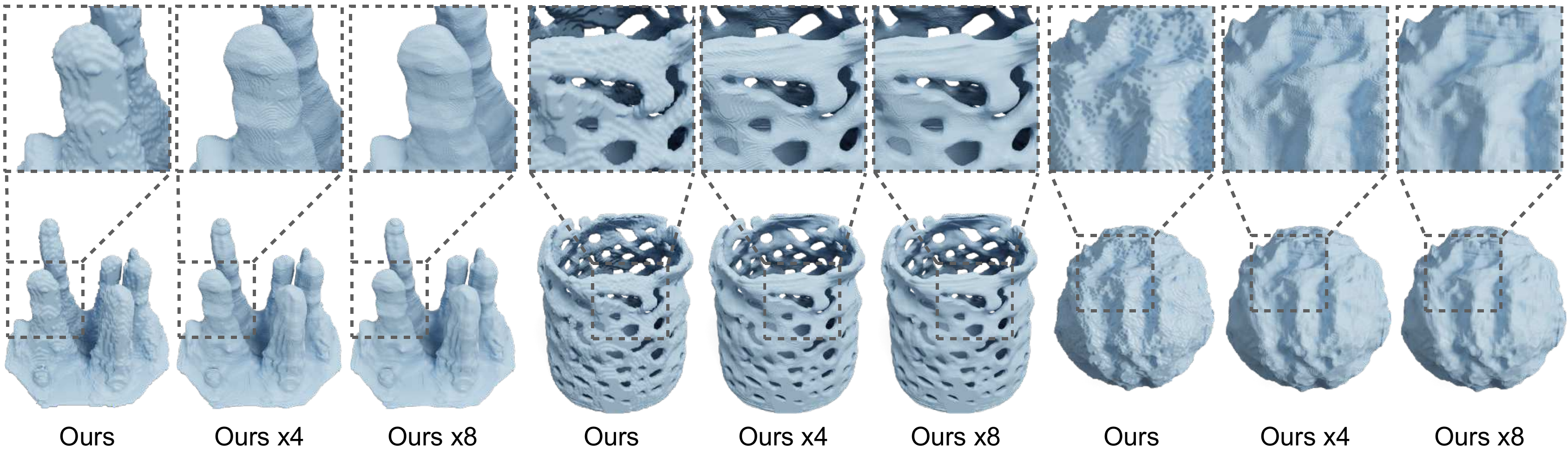}
  \caption{\textbf{Querying at a higher resolution.} 
  	We show $\times 4$/$\times 8$ upsampling results by querying $o(p)$ 
  	(resulted from the generated tri-plane representation)
    on a voxel grid with a $\times 4$/$\times 8$ training resolution.
  	The training resolutions for the three shape example are $128\times 100\times 106$, $128\times 128\times 128$ and $128\times 128\times 128$, respectively. In this figure, visualized meshes are the direct output from Marching Cubes \cite{marchingcubes}.
  }
  \label{fig:highres2}
\end{figure*}

We show more of our random generation results in \figref{gallery2} and \figref{gallery3}, more shape interpolation results in~\figref{interpolation2}, and the remaining comparison results in \figref{full_comp2}.
We also show more examples of querying higher resolution in \figref{highres2}.
In addition, we present detailed metric values for our ablation study in \tabref{full_ablation}.
Finally, we recommend the reader to open the offline webpage in the supplementary material, for interactive view of some generation and comparison results.

\section{Evaluation Metrics}
\label{sec:supp_metric}
For LP-IoU and LP-F-score, we sample voxel patches of size $11\times 11\times 11$ with a stride of $5$.
A patch is considered valid only if it has at least one occupied voxel and one unoccupied voxel in its center $3\times 3\times 3$ area.
We use all sampled valid patches from real shape as reference, denoted as $P_r$, and randomly select $1000$ sampled valid patches from the generated shape, denoted as $P_g$.
LP-IoU and LP-F-score are then defined as
\begin{equation}
	\text{LP-Score}(P_r, P_g)=\frac{1}{|P_g|}\sum_{x\in P_g} \mathbbm{1}[\max_{y\in P_r}\text{Score}(x, y) > \delta],
\end{equation}
where $\text{Score}$ is either IoU or F-score. 
F-score is calculated in a voxel-wise binary classification manner.
The threshold $\delta$ is set to $0.95$.

For SSFID, we take the 3D CNN classification network pretrained on ShapeNet \cite{chen2021decor} and use the deep feature map output by the second convolutional block to calculate the Fr\'echet Inception Distance (FID).
The deep feature map is spatially $4$ times smaller than the input voxel grid, and has $64$ channels.

We compute the diversity score in \figref{comparison} as follows.
For each shape in the a set of $k=100$ generated shapes $\{S_i\}$, we calculate its average distance to the other $k-1$ shapes.
Here we use $1-\text{IoU}(\cdot, \cdot)$ as the distance measure.
Then the score is the average value over all shapes in the set.
Concretely, the diversity score is defined as 
\begin{equation}
	\text{Div}(\{S_i\}) = \frac{1}{k}\sum_{1\le i \le k}[\frac{1}{k-1}\sum_{\substack{1\le j \le k \\ j\neq i}} 1- \text{IoU}(S_i, S_j)].
\end{equation}

\begin{table}[t]
\begin{tabular}{lllc}
\toprule
Figure    					    & Data          & Resolution                   & \makecell{\#Training \\ scales}\\
\midrule
\multirow{8}{*}{\makecell[l]{\figref{teaser} \& \\ \figref{gallery} \& \\ \figref{interpolation}}}
                                & Acropolis      & $256\times 72\times 118$      & 8 \\
                                & Vase   & $192\times 192\times 192$     & 8 \\
                                & Terrain        & $224\times 256\times 70$      & 8 \\
                                & Table  & $256\times 108\times 92$      & 8 \\
                                & Castle         & $256\times 70\times 242$      & 8 \\
                                & Desert cactus  & $110\times 256\times 120$     & 9 \\
                                & Stalagmites   & $256\times 200\times 212$     & 9 \\
                                & Rock stairs    & $128\times 36\times 256$      & 8 \\
\midrule
\multirow{8}{*}{\makecell[l]{\figref{gallery2} \& \\ \figref{interpolation}}}
                                & Canyon         & $256\times 256\times 50$      & 8 \\
                                & Plant pot      & $256\times 94\times 78$       & 8 \\
                                & Icelandic mountain     & $256\times 62\times 256$      & 9 \\
                                & Floating wood  & $256\times 70\times 84$       & 8 \\
                                & Cube hotel     & $146\times 256\times 126$     & 8 \\
                                & Log    & $66\times 256\times 112$      & 8 \\
                                & Boulder stone  & $256\times 80\times 108$      & 8 \\
                                & Small town     & $256\times 92\times 168$      & 8 \\
\midrule
\multirow{7}{*}{\makecell[l]{\figref{gallery3}}}
                                & Terrain 2      & $256\times 60\times 256$      & 9 \\
                                & Curved vase    & $162\times 148\times 192$     & 8 \\
                                & Elm tree       & $164\times 256\times 152$     & 9 \\
                                & Stone wall     & $256\times 56\times 138$      & 8 \\
                                & Ruined building	& $104\times 82\times 128$      & 6 \\
                                & Natural arch   & $132\times 142\times 256$     & 9 \\
                                & Industry house         & $120\times 98\times 256$      & 8 \\
\midrule
\multirow{10}{*}{\makecell[l]{\figref{comparison} \& \\ \figref{highres} \& \\ \figref{full_comp2}}}
                                & Acropolis      & $128\times 36\times 60$       & 6 \\
                                & Terrain        & $112\times 128\times 36$      & 6 \\
                                & Stalagmites    & $128\times 100\times 106$     & 7 \\
                                & Stone stairs   & $112\times 52\times 128$      & 7 \\
                                & Rock   & $128\times 128\times 128$     & 7 \\
                                & Wall   & $128\times 88\times 46$       & 6 \\
                                & Vase   & $128\times 128\times 128$     & 7 \\
                                & Cheese         & $48\times 78\times 128$       & 6 \\
                                & Cactus         & $74\times 128\times 72$       & 6 \\
                                & Tree   & $60\times 128\times 66$       & 6 \\
                               
\midrule
\figref{training_scales}
                                & Mountain with water      & $256\times 54\times 256$      & 9 \\
\midrule
\multirow{3}{*}{\makecell[l]{\figref{failure} \& \\ \figref{bias}}}
                                & Chair  & $104\times 128\times 100$     & 7 \\
                                & Tree branches  & $110\times 128\times 96$      & 7 \\
                                & Strange terrain        & $128\times 128\times 40$      & 7 \\
\bottomrule
\end{tabular}

\caption{\textbf{Data configuration.} 
}
\label{tab:data_config}
\end{table}

\begin{table*}[]
\begin{tabular}{llccccccccccl}
\toprule
\multirow{2}{*}{Metrics}    & \multirow{2}{*}{Methods} 		 & \multicolumn{11}{c}{Examples}                                                                   \\
&                          & Acropolis 		& Terrain 		 & Stalagmites 	  & Stairs 			& Rock  		 & Wall  			& Vase  		& Cheese 		& Cactus 		& Tree  		 & Avg. \\
\midrule
\multirow{4}{*}{LP-IoU
			$\uparrow$}     
& Proposed Method          & \textbf{0.902} & 0.647   		 & \textbf{0.254} & 0.777  			& \textbf{0.004} & 0.848 		 	& 0.175 		& 0.564  		& \textbf{0.583}& \textbf{0.015} & \textbf{0.477} \\
& Initial 2D Noises        & 0.816	   		& 0.638   		 & 0.247	   		  & 0.711			& 0.003			 & 0.818			& 0.174			& \textbf{0.579}& 0.412  		& 0.003 		 & 0.440 \\
& No Gaussian Blur         & 0.887	   		& \textbf{0.648}	 & 0.249	   		  & 0.756			& 0.002			 & \textbf{0.854}	& 0.173			& 0.561	 		& 0.407  		& 0.003 		 & 0.454 \\
& No Weights Reuse         & 0.874	   		& 0.637	 		 & 0.220	   		  & \textbf{0.821}	& \textbf{0.004} & 0.821			& \textbf{0.179}& 0.575	 		& 0.374  		& 0.005 		 & 0.451 \\
\midrule
\multirow{4}{*}{LP-F-score
			$\uparrow$} 		
& Proposed Method          & \textbf{0.950} & \textbf{0.812} & \textbf{0.363} & 0.865  			& \textbf{0.129} & \textbf{0.892} 	& 0.214 		& \textbf{0.716}& \textbf{0.847}& \textbf{0.149} & \textbf{0.594} \\
& Initial 2D Noises        & 0.884	   		& 0.809	 		 & 0.360	   		  & 0.803			& 0.112			 & 0.867			& 0.212			& 0.715	 		& 0.730  		& 0.074 		 & 0.557 \\
& No Gaussian Blur         & 0.933	   		& 0.811	 		 & 0.356	   		  & 0.854			& 0.103			 & 0.889			& 0.215			& 0.687	 		& 0.716  		& 0.065 		 & 0.563 \\
& No Weights Reuse         & 0.931	   		& 0.793	 		 & 0.316	   		  & \textbf{0.929}	& 0.112			 & 0.864			& \textbf{0.223}& 0.705	 		& 0.720  		& 0.085 		 & 0.568 \\
\midrule
\multirow{4}{*}{SSFID
			$\downarrow$}   
& Proposed Method          & \textbf{0.037} & \textbf{0.050} & \textbf{0.078} & 0.102  			& \textbf{0.020} & \textbf{0.272} 	& \textbf{0.029}& 0.065  		& \textbf{0.018}& \textbf{0.073} & \textbf{0.074} \\
& Initial 2D Noises        & 0.058	   		& 0.051	 		 & 0.081	   		  & 0.070			& 0.022			 & 0.376			& 0.033			& \textbf{0.064}& 0.028  		& 0.089 		 & 0.087 \\
& No Gaussian Blur         & 0.035	   		& 0.060	 		 & 0.087	   		  & 0.048			& 0.025			 & 0.378			& 0.036			& 0.088	 		& 0.032  		& 0.093 		 & 0.088 \\
& No Weights Reuse         & 0.039	   		& 0.056	 		 & 0.116	   		  & \textbf{0.032}	& 0.027			 & 0.303			& 0.033			& 0.095	 		& 0.033  		& 0.086 		 & 0.082 \\
\bottomrule
\end{tabular}
\caption{\textbf{Ablation study.} $\uparrow$: a higher metric value is better; $\downarrow$: a lower metric value is better. See \secref{ablation} for the description of each variant.}
\label{tab:full_ablation}
\end{table*}

\end{document}